\documentclass[12pt]{article} 
\usepackage{abstract}
\usepackage[utf8]{inputenc} 
\usepackage{amsxtra} 
\usepackage{amsthm}
\usepackage{amssymb}
\usepackage{amsfonts}
\usepackage{graphicx}
\usepackage{rotating}
\usepackage{color}
\usepackage{relsize}
\usepackage{epsfig}
\usepackage{subcaption} 
\usepackage{verbatim}
\usepackage[printonlyused]{acronym} 
\usepackage{setspace}
\usepackage{epstopdf}
\usepackage{authblk}
\usepackage{sectsty}
\usepackage[square,numbers]{natbib}
\usepackage[colorinlistoftodos]{todonotes}
\usepackage[colorlinks=true,allcolors=black]{hyperref}
\usepackage{textcomp,marvosym}
\usepackage{soul}
\usepackage{algorithm}
\usepackage{algorithmicx}
\usepackage{algpseudocode}
\usepackage{url}
\usepackage{hyperref}
\usetikzlibrary{shapes,arrows,automata,positioning}
\usepackage{mathtools}
\usepackage{dsfont}

\usepackage{geometry} 
\usepackage{graphicx} 
\usepackage{amssymb}
\usepackage{grffile} 
\usepackage{natbib}
\usepackage{graphics}
\usepackage{epsfig} 
\usepackage{dcolumn} 
\usepackage{bm} 
\usepackage{hyperref} 
\usepackage{wrapfig}
\usepackage{framed}
\usepackage{amsmath}
\usepackage{tabularx} 
\usepackage{caption}
\usepackage{float} 
\usepackage{soul}
\usepackage{physics}
\usepackage{etaremune}
\usepackage{enumitem}
\usepackage[final]{pdfpages}
\usepackage{xcolor}
\usepackage{xspace}
\usepackage{accents}
\usepackage{url,comment}
\graphicspath{{./figures/}}

\renewcommand{\l}{\left}
\renewcommand{\r}{\right}

\begin{document}


\title{Lagrangian Large Eddy Simulations\\ via Physics Informed Machine Learning}

\author[a]{Yifeng Tian}
\author[b,c]{Michael Woodward} 
\author[b]{Mikhail Stepanov}
\author[c]{Chris Fryer}
\author[b,c]{Criston Hyett}
\author[c]{Daniel Livescu}
\author[b]{Michael Chertkov}

\affil[a]{Information Sciences Group, Computer, Computational and Statistical Sciences Division (CCS-3), Los Alamos National Laboratory, Los Alamos, NM, 87545, USA.}

\affil[b]{Graduate Interdisciplinary Program in Applied Mathematics and Department of Mathematics, University of Arizona, Tucson, AZ, 85721, USA.}
\affil[c]{Computational Physics and Methods Group, Computer, Computational and Statistical Sciences Division (CCS-2), Los Alamos National Laboratory, Los Alamos, NM, 87545, USA.}




\maketitle
\begin{abstract}
High Reynolds Homogeneous Isotropic Turbulence is fully described within the Navier-Stokes (NS) equations, which are notoriously difficult to solve numerically. Engineers, interested primarily in describing turbulence at a reduced range of resolved scales, have designed heuristics, known as Large Eddy Simulation (LES). LES is described in terms of the temporally evolving Eulerian velocity field defined over a spatial grid with the mean-spacing correspondent to the resolved scale. This classic Eulerian LES depends on assumptions about effects of sub-grid scales on the resolved scales. 

Here, we take an alternative approach and design novel LES heuristics stated in terms of Lagrangian particles moving with the flow. Our \emph{Lagrangian LES}, thus L-LES, is described by equations generalizing the weakly compressible Smoothed Particle Hydrodynamics formulation with extended parametric and functional freedom, which is then resolved via Machine Learning training on Lagrangian data from Direct Numerical Simulations of the NS equations. The L-LES model includes physics-informed parameterization and functional form, by combining physics-based parameters and physics-inspired Neural Networks to describe the evolution of turbulence within the resolved range of scales. The sub-grid scale contributions are modeled separately with physical constraints to account for the effects from un-resolved scales. We build the resulting model under the Differentiable Programming framework to facilitate efficient training. We experiment with loss functions of different types, including physics-informed ones accounting for statistics of Lagrangian particles. We show that our Lagrangian LES model is capable of reproducing Eulerian and unique Lagrangian turbulence structures and statistics over a range of turbulent Mach numbers.
\vspace{6pt} \vspace{6pt}
\end{abstract}

Accurate Direct Numerical Simulations (DNS) of turbulent flows in physical sciences and engineering applications are, in general, prohibitively expensive due to the existence of a wide range of length and time scales. This challenge has motivated the development of the Reduced Order Models (ROM) which achieve fast and efficient solutions in practical applications. In the field of turbulence simulations, Reynolds-Averaged Navier-Stokes (RANS) and Large Eddy Simulations (LES) have been widely adopted as alternatives for the DNS of practical turbulent flows \cite{sagaut2006large}. In these traditional methods, turbulent flows are viewed from the Eulerian frame where the computational domain is discretized into small elements or units via meshes. In LES, a spatio-temporal low-pass filter is applied to the Navier-Stokes (NS) equations, therefore removing small-scale fluctuations from the consideration. LES results in the reduction of the computational complexity, simply because the range of scales that need to be resolved \cite{sagaut2006large} is reduced. 

Eulerian LES has achieved great success in scientific and engineering applications. However, when solving turbulent problems with complex geometries, moving interfaces, and multi-phase materials, the Eulerian description suffers from deteriorating numerical accuracy due to deforming mesh, fluid-particle coupling, etc. To resolve these issues, resulted from Eulerian description of the flow, mesh-free methods have been proposed \cite{liu2005introduction,liu2010smoothed}, such as Smoothed Particle Hydrodynamics, vortex method, and mesh-free Galerkin method. Simulating turbulent flows using the mesh-free, Lagrangian approximation of Navier-Stokes equations is appealing because it provides a unique perspective for understanding the transport processes, such as the mixing and dispersion of passive scalars. Such processes are dominated by the advective motion of velocity fluctuations in time and space, which is naturally a part of the Lagrangian representation of the fluid via trajectories of representative fluid particles \cite{yeung2002lagrangian}. Among the mesh-free methods for solving fluids problems, Smoothed Particle Hydrodynamics (SPH) \cite{gingold77,monaghan1992smoothed,monaghan2012smoothed} has been  adopted in a wide range of scientific applications, e.g. astrophysics, computer graphics, free-surface flows, fluid-structure interaction, bio-fluids, geological flows, magnetohydrodynamics \cite{1994ApJ...435..339H,1999Icar..142....5B,1995MNRAS.277..362B,1999A&A...341..499R,1999ApJ...525..554F,1999ApJ...527L...5B,2005MNRAS.364.1105S,2006MNRAS.373.1074S,2012JCoPh.231..759P,shadloo2016smoothed,lind2020review,ye2019smoothed,2022arXiv220105896R}. Recently, there is a number of studies adopting SPH to model turbulence \cite{2002MNRAS.335..843M, lo2002simulation, dalrymple2006numerical, price2012resolving, mayrhofer2015dns,2018PASA...35...31P,2021MNRAS.506.2836R}. SPH framework was also integrated into RANS modeling, and specifically into RANS two-equation models, such as the $k-\epsilon$ type \cite{violeau2007numerical,de2016sph,leroy2014unified}. A series of studies explored similarity of the filtering procedure of LES and the smoothing procedure of SPH to develop LES models in the SPH framework \cite{bicknell1991equations,di2017smoothed,antuono2021smoothed}. These papers have helped to establish SPH as a promising framework for modeling turbulence which accounts for advection explicitly,  though integration over Lagrangian trajectories. However, the papers have also highlighted significant challenges of the SPH modeling related to resolving the stress tensor term in the Lagrangian frame and also the multi-particle nature of the formulation, making derivation and tuning parameters within the SPH models time-consuming and difficult \cite{di2017smoothed}. 

This manuscript addresses the challenge of adopting the Lagrangian approach, and specifically generalizing,  i.e adApting -- NOT adOpting, the SPH approach, to reduce-order modeling of NS turbulence directly. Here, we mention several works that attempt to adapt the SPH approach based on different numerical procedures, such as mesh-less finite volume \cite{hopkins2015new} and adaptive smoothing kernel \cite{owen1998adaptive} and differentiate them from the more general formulation proposed in the work, with physics-informed and interpretable parameterization that can be learned from data. We achieve this goal by utilizing the power of Machine Learning (ML),  and more generally of the Artificial Intelligence (AI).

Modern ML and specifically Deep Learning, leveraging efficient computational tools such as automatic differentiation and sensitivity analysis of forward and backward propagation, has resulted in multiple success stories in classic AI disciplines, such as computer vision \cite{sebe2005machine}, speech recognition \cite{deng2013machine} and natural language processing \cite{goodfellow2016deep}. However, the biggest achievements of AI and ML in sciences so far had been limited, until recently, to approaches that are data-driven but agnostic to traditional scientific modeling of the underlying physics. Integrating the breakthrough of the modern AI and ML with physical modeling,  and specifically for this manuscript with Lagrangian modeling of turbulence,  is the major challenge of what we call today Physics-Informed Machine Learning (PIML) \footnote{It may be appropriate to mention here that PIML, as a term, was coined in 2016 in the name of the LANL workshop in Santa Fe, NM, which later became a bi-annual event with the 4-th PIML taken place in May of 2022 \cite{PIML}. The workshops brought together ML/AI experts and statistical physicists to present their work where either principal ideas from statistical physics were utilized in the design and training of ML models, including Neural Networks,  or ML techniques were integrated into physical models to better describe applications in various quantitative disciplines of sciences and engineering. However, the term was also used later by the authors of \cite{Karniadakis2021Physic-Informed} to describe Neural Network modification of the classic collocation method (for the numerical solution of ODEs and PDEs, see \cite{lapidus2011numerical}),  originally introduced in \cite{Lagaris1998Artificial} and then popularized in \cite{Raissi2019Physics-informed} as the Physics Informed Neural Networks (PINN).}. In the field of turbulence modeling, there has been a surge of PIML activity in recent years. In particular, major efforts have been devoted to the development of closure models for Reynolds Averaged Navier-Stokes (RANS) and Large Eddy Simulations (LES) in the Eulerian frame using innovative Neural Network (NN) architectures. Some of the contributions, prioritized here due to their significance for the methodology of this manuscript, are:  Tensor Basis Neural Network (TBNN) embedding physical constraints, such as Galilean invariance and rotational invariance, into the closure model \cite{ling2016reynolds}; and  PIML models infusing physical constraints into the neural networks \cite{2018PIML-LANL,wang2017physics,mohan2020div}.
A comprehensive overview of these and other related contributions to the field of turbulence closure modeling in Eulerian frame can be found in  \cite{duraisamy2019turbulence}. Other than developing closure models for RANS and LES, researchers have been experimenting with novel ML approaches to learn turbulence dynamics. In this regard, and just to name a few, we mention \cite{mohan2019compressed,mohan2020jot}, where a Convolutional Long Short Term Memory (ConvLSTM) Neural Network was developed to learn spatial-temporal turbulence dynamics; studies of super-resolution allowing to reconstruct turbulence field using under-resolved data \cite{fukami2019super}; and Neural Ordinary Differential Equation (Neural ODE) for turbulence forecasting \cite{portwood2019turbulence}. PIML models for Lagrangian description of turbulence have also been actively developed. Authors of \cite{ladicky2015data} have used a regression forest to approximate the behavior of particles in Lagrangian fluid simulations. A differentiable fluid solver accounting for Lagrangian dynamics has been developed in Ref. \cite{schenck2018spnets}. A continuous convolution network was constructed to learn SPH-inspired Lagrangian simulation in \cite{ummenhofer2019lagrangian}. Three of us (YT, DL, and MC) have reported in \cite{tian2021physics} development of a PIML approach to design a closure model for the Lagrangian (single particle) dynamics of the Velocity Gradient Tensor (VGT). 

We have also developed in \cite{woodward2021physics} a hierarchy of reduced Lagrangian multi-particle models of the PIML/SPH type trained on ground truth data of two types, as described below. First, we train the SPH model on synthetic multi-particle simulations imitating multi-scale Lagrangian turbulence.  Then, we also extend it to the Lagrangian ground truth data derived from the fully-resolved Eulerian DNS of homogeneous isotropic turbulence.

In this manuscript, we further develop the ideas from the Lagrangian modeling of the PIML/SPH type we have started to explore in \cite{woodward2021physics}.  In  \cite{woodward2021physics}, we took advantage of the degrees of freedom in the definition of SPH and then ad{\bf O}pted SPH to the ground truth data,  i.e. adjusted the degrees of freedom within SPH to fit the data. On the contrary, in this manuscript we generalize SPH,  i.e. ad{\bf A}pt SPH. This generalization allows us to
introduce a general Lagrangian LES (L-LES) model, simulating turbulence at the resolved scales, and to train this model on the Lagrangian data derived from the fully-resolved Eulerian DNS.

We describe the L-LES framework in the remainder of the manuscript in steps. The stage is set in Section \ref{sec:llesmodel} where we introduce (fictitious) particles, and describe their Lagrangian dynamics, i.e. effective advection by a collective velocity field, and effective interaction imposed by the collective contribution of the particles to the inter-particle forces, reconstructed from the velocity field. Parametric, as well as functional, degrees of freedom in the resulting non-linear system of multi-particle ODEs, stated in terms of coordinates and velocities of the particles, are discussed both in intuitive physical but also formal mathematical terms. We argue about why and in which sense the system of equations governing dynamics of interacting Lagrangian particles is a good LES heuristics. In Section \ref{sec:kernel}, the connection between the Lagrangian and Eulerian descriptions of turbulence is established through Smoothing Kernel (SK). We introduce parameterization of the SK via a Neural Network (NN), and then set an SK- Loss Function (LF), based on the turbulence statistics (connecting both Lagrangian and Eulerian descriptions) to train the SK-NN.  
We provide a comprehensive description of how we "learn" the PIML / L-LES model in Section \ref{sec:piml}. We discuss the choice of the L-LES loss function, respective L-LES NN architecture, and relevant technical details. 
Section \ref{sec:gtdata} is devoted to details of the training and validation experiments.  We explain the ground truth DNS model, process of data generation, and significance of validating/ testing the underlying physics in detail. We summarize results and discuss path forward, respectively, in Section \ref{sec:results} and Section \ref{sec:conclusions}.

\section{Lagrangian Large Eddy Simulation}

\label{sec:llesmodel}

In this Section we describe the main principles of the proposed Lagrangian Large Eddy Simulation (L-LES) methodology for modeling turbulent flows using a system of interacting Lagrangian particles, whose evolution is governed by a set of parameterized equations with embedded physical constraints.

In the classic mesh-free Lagrangian models of turbulence \cite{liu2005introduction,liu2010smoothed}, the dynamics of acceleration, density, and pressure fields evaluated at the particle positions, are approximated following the Lagrangian version of the Navier-Stokes equations. This system of interacting particles provides an excellent platform for simulating turbulent flows in the Lagrangian frame, especially for resolving large-scale structures in the context of Large Eddy Simulation (LES). This is because dynamics of a Lagrangian particle cloud represent accurately sweeping of an eddy associated with the mean particle distance by larger eddies. The interactions among particles in the cloud span over a wide range of scales, which enables the system to capture the multi-scale turbulence dynamics even at the scales which are smaller than the filtering length (sub-grid scale) because the distances between non-uniform Lagrangian particles may become smaller than the filtering length. However, to formally derive a set of dynamical equations that describe Lagrangian particles in the spatially-filtered turbulence field, additional closure models for the emerged sub-filter contributions are needed. In the classic LES, which are to the best of our knowledge exclusively Eulerian, these closure models have been historically formulated  based on heuristics and hand-tuned parameters. Very recently, the PIML framework has been started to be used to improve the parameter tuning, or even to discover new functional forms of the closure for Eulerian LES \cite{duraisamy2019turbulence, portwood2021interpreting}. In this study, we aim to leverage the PIML framework in the Lagrangian, multi-particle framework to learn Lagrangian LES (L-LES) models from filtered DNS data, which alleviates us from the exhaustive process of model tuning. 

\subsection{Pair-Wise Particle Interaction}\label{sec:pair-wise}

A straightforward phenomenological model for the evolution of a system of  Lagrangian particles is through the interactions among the particles. In this work, we utilize the main physics idea of the particle-based methodology -- i.e., express the flow, e.g., velocity and density fields evolving in time -- in terms of particles. Specifically, we consider particles "interacting" with each other in a pair-wise way, where the force exerted on a given particle by the flow is expressed as a sum of terms, each dependent on the current position of the particle and another particle (thus a pair),  but also dependent on the velocities and densities evaluated at positions of the two particles. The pair-wise expression for the force acting on any particle in the volume is common for all Particle-Based Methods (PBM), including arguably the most popular example of the PBM -- the Smoothed Particle Hydrodynamics. Other PBM examples include the family of the mesh-free Galerkin methods and  Molecular Dynamics (MD) methods, see   \cite{liu2003smoothed,liu2005introduction,liu2010smoothed} and references therein. 

Notice that the prime use of the particle-based, Lagrangian methods, is for either description of sub-viscous scale phenomena in materials (largely MD methods) and flows or for descriptions of the large scale (energy-containing) part of the flows. In the sub-viscous setting, particles are associated with the actual physical particles or small patches of flows resolved at the scales which are of the order or smaller than the viscous (Kolmogorov) scale of the actual flow. In the case of large-scale flow modeling, the particle-based methods (primarily SPH) have been used for a sketchy (imprecise) description of the underlying physics at the largest (often astronomical) scales. The main goal of this manuscript is to bridge the gap between the smallest (viscous) and the largest (energy-containing) scales in a continuum description of the flow. We focus here on building L-LES by resolving the range of scales that covers a significant portion of the large-scale part of the inertial range of scales in a turbulent flow. 

The traditional approach of the particle-based methods describing continuum level flows consists of an intuitive derivation of the inter-particle interaction by analogy with the Navier-Stokes equations. In this work, we choose to generalize the approach.  We do not derive the interactions by analogy with the Navier-Stokes equations, but rather model them in the most general way.  This general way of modeling is inclusive in the sense that it allows us to include many other modeling ideas used in the LES and particle-based methods so far.

For an $N$ particles system in three-dimensional domain $\Omega\subset \mathbb{R}^3$, we introduce a vector, $\bm{\phi}_i=[\bm{v}_i, \rho_i]^T$, built from velocity and density fields evaluated at the particle locations,  $\bm{x}_i \in \Omega$.  (Given that our focus here is on the weekly compressible limit of fully compressible flows, we do not account for explicit dependence on the pressure field, assumed expressed via density according to a barotropic equations of state.) Then, the most general equations we consider describing the evolution of the system of particles are:
\begin{eqnarray}
\frac{d}{dt} \begin{bmatrix} \bm{x}_i \\ \bm{\phi}_i \end{bmatrix} &=& \begin{bmatrix} \bm{v}_i  \\ \displaystyle\sum_{j=1, j\neq i}^{N} \bm{f}(\bm{x}_{i}, \bm{x}_j, \bm{\phi}_i, \bm{\phi}_j) \end{bmatrix}, \, \forall \, i \in 1,...,N, \label{eqn:pairparticle}
\end{eqnarray}
where $\bm{f}$ is (yet to be learned/discovered) function expressing the pair-wise force imposed on a particle $i$ by a particle $j$ and dependent on the positions of the two functions as well as the vector $\bm{\phi}$ evaluated at the positions of the two particles.

In general, we have the freedom to choose the functional form of $\bm{f}$ in Eq.~(\ref{eqn:pairparticle}) for modeling, by either using heuristic physical/mathematical arguments or inferring it from data. For example, in the specific choice made within the context of Smoothed Particle Hydrodynamic modeling, $\bm{f}$ is modeled by approximating the right-hand side (RHS) of the Lagrangian NS equations using the reconstructed field by a parameterized smoothing kernel. SPH is formulated to ensure that the physical constraints, such as conservation of momentum, Galilean and rotational invariance, are satisfied. Readers interested in the details on the SPH modeling and learning of NS turbulence are advised to check \cite{woodward2021physics}. In this manuscript, we do account for the relevant physical constraints, however, we do not follow the PIML/SPH learning path of \cite{woodward2021physics}. Instead, we take here a more general approach, described in the following. 

\subsection{Physics-Informed Choice of the Pair-Wise Interaction}

Despite the existence of a wide range of functions that might describe the data, a Neural Network, as a general function approximator with over-parameterization, provides an excellent data-driven platform for modeling pair-particle interaction. In the PIML framework, which aims to leverage the capabilities of NNs in advancing scientific applications, the mathematical and physical laws and constraints are either built into the architectures of NN or integrated into the loss functions. Various studies have shown that the PIML framework is advantageous in improving robustness, efficiency, and interpretability over a physics-blind one \cite{mohan2020jot, tian2021physics}. In this Section, we describe how the physics constraints are included in (otherwise  generic) pair-wise function $\bm{f}$ in Eq.~(\ref{eqn:pairparticle}).

We need to enforce invariance of the dynamics (\ref{eqn:pairparticle}) under the Galilean, translational, and rotational transformations of the underlying frame of reference. Let us denote the operator, representing an element of the union of the physical transformations, as $\mathcal{T}$. Then, we require that the pair-wise force is $\mathcal{T}$-invariant, i.e. formally,
\begin{equation}
    \mathcal{T}f(\bm{x}_i, \bm{x}_j, \bm{\phi}_i, \bm{\phi}_j ) = \bm{f}(\mathcal{T}\bm{x}_i, \mathcal{T}\bm{x}_j, \mathcal{T}\bm{\phi}_i, \mathcal{T}\bm{\phi}_j). 
\end{equation}

If $\mathcal{T}$ is limited to the translational invariance, then transformation of $\bm{x}_i$ can be expressed as $\mathcal{T}\bm{x}_i = \bm{x_i} +\bm{x}'$, whereas $\phi_i$ stays invariant, i.e. $\mathcal{T}\bm{\phi}_i = \bm{\phi_i}$. Therefore, a straightforward approach to enforce translational invariance of $f$ is to replace the general dependence on $\bm{x}_i$ and $\bm{x}_j$ on dependence on only the difference of the two, $\bm{x}_i-\bm{x}_j$. Similarly, the Galilean invariance, which is associated with the transformation to a different inertial frame, $\mathcal{T}\bm{v}_i=\bm{v}_i + \bm{v}', \mathcal{T}\bm{x}_i=\bm{x}_i + \bm{v}'t$, can be enforced by replacing the general dependence of the two particle velocities, $\bm{v}_i$ and $\bm{v}_j$, by the relative velocity, $\bm{v}_i - \bm{v}_j$. To enforce the rotational invariance, symmetry transformations of the scalar (density) and the vector (velocity) components of $\phi$, $\bm{v}_i$, should be dealt with differently. Let us decompose $\bm{f}$ into its scalar, $f_s$, and  vector, $\bm{f}_v$, components. Then rotational invariance of the scalar component of the force requires that, $\mathcal{T}f_s = f_s$. The respective constraint which needs to be imposed on the transformed vector component of $\bm{f}_v$, $\mathcal{T}\bm{f}_v$, is that it becomes a linear combination of the rotation-aware vector bases and scalar functions of rotational invariant scalars (we follow here the so-called Tensor Basis (TB) approach of \cite{ling2016reynolds,tian2021physics}).

Another important aspect in developing our PIML model is the capability of generalizing over a wider range of parameters that have not been used in the model training. In the context of weakly compressible Navier-Stokes turbulence, we are interested in developing a Lagrangian LES model that can describe turbulence with different turbulent Mach numbers, $M_t$~\footnote{$M_t$ is the turbulent Mach number, which is estimated as the ratio of the typical velocity at the energy-containing scale to the speed of sound, $M_t=v_L/c_s$. In the limit of low Mach number, for single component flows, density can be expanded as $\rho \approx \rho_0 + \rho_1$, where $\rho_1 \sim M^2_t \rho_0$ \cite{LivescuARFM}, so that $M_t^2$ is proportional to the fluid density deviation from the uniform distribution. Note that any particle-based modeling of turbulence requires introducing, discussing, and analyzing compressibility simply because any distribution of particles translates into fluid density which is always spatially non-uniform, even if slightly. Therefore, even if we model fully incompressible turbulence, we should still introduce an effective turbulent Mach number when discussing a particle-based approximation, as $M_t^2 \sim |\rho-\rho_0|/\rho_0$.}, at fixed Reynolds number~\footnote{$Re$ number measures a relative strength of the self-advection term to the viscous term evaluated at the integral scale of turbulence. It is thus estimated as $Re=v_L L/\nu=\varepsilon^{1/3}L^{4/3}/\nu$ \cite{Frisch1995Turbulence}. However, as a part of G.I. Taylor legacy, we often measure the strength of turbulence in terms of the so-called, $Re_\lambda= v_L \lambda/\nu$, where $\lambda=L \sqrt{15/Re}$ is the so-called Taylor micro-scale, which is sandwiched in between the energy-containing scale, $L$, and the Kolmogorov (viscous) scale, $\eta$, i.e. $\eta<\lambda<L$ \cite{Tennekes1978AFirst}.}. 

Our next key step is to leverage the enormous progress achieved during recent decade in using NNs as universal function approximators and represent the remaining freedom left in the functions on the right-hand-side of Eqs.~(\ref{eqn:pairparticle}) via NNs. This path leads us to the following NN version of Eqs.~(\ref{eqn:pairparticle}) consistent with the physical symmetries (expressed according to the TB approach):
\begin{eqnarray}
\frac{d\bm{\phi}_{i}}{dt}=\frac{d}{dt} \begin{bmatrix} \rho_i \\ \bm{v}_i \end{bmatrix} &=&
\sum_{j=1}^{N} \begin{bmatrix} \mathcal{NN}_\rho (I_{ij,m}, m \in 1,...,5; \lambda_\rho )\\ \displaystyle \sum_{k=1}^{2} \mathcal{NN}_{v,k}(I_{ij,m}, m \in 1,...,5; \lambda_v ) \bm{b}_{ij,k}+\Pi_{ij}\bm{b}_{ij,1} \end{bmatrix} + \begin{bmatrix} 0 \\F_i \end{bmatrix}, \label{eqn:nnmodel}
\end{eqnarray}
where $N$ is the number of particles placed inside of the computational domain, $\Omega$. In all the numerical experiments, reported in this manuscript, $\Omega$ is a  three dimensional cube of the volume, $(2\pi)^3$, i.e. $\Omega=[0,2\pi]^3$. Here in Eq.~(\ref{eqn:nnmodel}) $I$ and $\bm{b}$ denote, respectively, the scalar invariants and vector bases of the bases expansion approach of \cite{ling2016reynolds,tian2021physics} and are defined using pair-particle information:
\begin{eqnarray*}
\bm{x}_{ij} &=& (\bm{x}_i - \bm{x}_j)/d, \, \bm{v}_{ij} = (\bm{v}_{i} - \bm{v}_j)/v_{rms}, \, \rho_{ij} = \frac{1}{2} (\rho_i +\rho_j)/\rho_{rms}, \\
I_{ij,1} &=& \rho_i/\rho_{rms}, \, I_{ij,2} = \rho_j/\rho_{rms}, \, I_{ij,3} = \vert \bm{x}_{ij} \vert, \, I_{ij,4} = \vert \bm{v}_{ij} \vert, \, I_{ij,5} = \bm{x}_{ij}\cdot\bm{v}_{ij}, \\
\bm{b}_{ij,1} &=& \bm{x}_{ij}, \, \bm{b}_{ij,2} = \bm{v}_{ij}.
\end{eqnarray*}

To ensure the generalizability of the L-LES model, we formulate it in non-dimensional form, as shown above. In this work, we use the filtered RMS velocity $v_{rms}$ and average pair-particle distance $d$ as the reference velocity and length scales, while the density fluctuations are normalized by $\rho_{rms}$. In particular, the latter makes the features independent of $M_t$ to leading order. The reference length scale $d$, which represents the resolved turbulence field, is chosen to be within the inertial subrange. With this normalization of features, we expect the learned dynamics can be generalized over different $d$ values within the inertial range, as this is assumed to be universal \cite{Tennekes1978AFirst}. On the other hand, the forcing term is used in the ground-truth data generation to control $M_t$ and $Re_{\lambda}$ and contains an explicit dependency on $M_t$. Thus, through the non-dimensionalization of the input features of NNs, we separate the modeling of pair-particle interaction into $M_t$ independent (represented by NN) and $M_t$ dependent (reference scales) parts. When generalizing over different $M_t$,  the same NN can be employed, while the overall magnitude of the acceleration learned from the ground truth data can be used to re-scale the NN output to the correct magnitude.

Let us discuss other details of the parameters and terms contributing to Eqs.~(\ref{eqn:nnmodel}):
\begin{itemize}
    \item \underline{Neural Networks:} We introduce two distinct NNs, $\mathcal{NN}_{\rho}$ and $\mathcal{NN}_{v}$, to approximate respective components  of the function $\bm{f}$. As customary,  the NNs will over-parameterize to fit the data. We will utilize in the following, specifically in Section \ref{sec:piml} where numerical validation of the L-LES approach is discussed, the same neural network for both $\mathcal{NN}_{\rho}$ and $\mathcal{NN}_{v}$ -- containing 5 scalar features as input and 4 hidden layers, each with 100 neurons. The output of $\mathcal{NN}_{\rho}$ is a scalar that accounts for the RHS of the continuity equation. The outputs of $\mathcal{NN}_{v}$ are the scalar coefficients of the two vector bases. 
    
    \item \underline{Eddy Viscosity:} We make sure that the eddy diffusivity, $\Pi$-dependent, term in Eq.~(\ref{eqn:nnmodel}) satisfies physical symmetries (see discussion above) and 
    otherwise we use the following form introduced in \cite{monaghan1985refined}, which has been used as the standard artificial viscosity in SPH literature.
\begin{eqnarray*}
\Pi_{ij} &= & \begin{cases}
\frac{-\alpha c\mu_{ij} + \beta \mu_{ij}^2}{\rho_{ij}}, & \bm{v}_{ij} \cdot \bm{x}_{ij} < 0,\\
0, & \bm{v}_{ij} \cdot \bm{x}_{ij} \geq 0
\end{cases} \\
\mu_{ij}&= &\frac{d\bm{v}_{ij}\cdot \bm{x}_{ij}}{\vert \bm{x}_{ij}\vert^2 + \epsilon d^2}, 
\end{eqnarray*}

The eddy viscosity term associated with a pair of particles activates (becomes significant) when the two particles are separated from each other at the distance which is smaller than the resolved scale, $d$. We naturally link the resolved scale, $d$, to the number of particles, $N$, thus setting $d=2\pi/N^{1/3}$ in our L-LES model over the $[0,2\pi]^3$ cube. Therefore the eddy-viscosity term leads, effectively, to repulsion elongated with $\bm{b}_{ij,1}$, and it can thus be interpreted (according to the name) as modeling effective energy transfer from the resolved scales to smaller scales. Also, and consistently with the common reasoning, well documented in the SPH literature (see e.g. \cite{ye2019smoothed}), the eddy viscosity term prevents particles from clustering and collisions, therefore, providing overall numerical stability of the scheme. The coefficients, $\{ \lambda_\rho, \lambda_v, \alpha, \beta\}$, contributing to the eddy viscosity term are not fixed (as custom in the classic particle-based method) but are learnable, i.e. subject to optimization together with the parameters of the aforementioned NN. The eddy viscosity term could be absorbed into the above-mentioned NN term since NN can be used as a universal function approximator, but here we separate the parameterization of the eddy viscosity term to inject physical intuition and mathematical structures. This may also alleviate the burdens on learning NN parameters so that the learning process becomes easier and smoother. 

\item \underline{External Force:} The last term $F_i$ in (\ref{eqn:nnmodel}) models the external force, which is large-scale, i.e. it injects energy into the system largely at the scales comparable to the size of the box, $L$. We choose the forcing term in the L-LES model (\ref{eqn:nnmodel}) to be linear in ${\bm v}_i$
\begin{gather}\label{eq:F}
F_i = \alpha_F\sum_{j}^{N}\chi_F(\bm{x}_i-\bm{x}_j)\bm{v}_i, 
\end{gather}
consistently with the structure of the forcing term used in our ``ground-truth'' DNS data to 
ensure that the DNS (and thus our L-LES model too) reaches a statistically-steady state.
Here in Eq.~(\ref{eq:F}), $\alpha_F$ is a parameter, which is set to match the energy injection rate from external forcing with the dissipation rate of the turbulent flow; and $\chi_F(\bm{x}_i-\bm{x}_j)$ is the forcing weight function, dependent on the inter-particle radius-vector in the form prescribed by the DNS. In all of our experiments the large-scale forcing term is fixed, i.e. it is the only term on the RHS of Eq.~(\ref{eqn:nnmodel}) which is not subject to fitting/learning. 

\end{itemize}

A number of comments about the future use of the L-LES model (\ref{eqn:nnmodel}) in the remainder of the manuscript are in order. 

First, it is important to emphasize that all the parameters contributing to the L-LES model are subject to learning, i.e. optimization minimizing the so-called Loss Function (LF) expressing the mismatch between DNS data and the L-LES model. There are many plausible and also physics-informed choices for the LF, e.g. based on Lagrangian trajectories, Eulerian field, and statistics, which we discuss in detail in Section \ref{sec:gtdata}. 

Second, let us recall that the L-LES model is set as a phenomenology/heuristics which is stated in terms of the particles that are not representing exactly the actual particles advected by the velocity fields from the DNS (our ground truth data). Indeed, we aim to construct L-LES as a statistical model -- i.e. model reproducing correctly only important statistical features of turbulence but not exact (deterministic) trajectories. To achieve this statistical goal, we will introduce in the next Section a mapping between Lagrangian particles and the Eulerian field, which will then be used to formulate the learning problem and then perform training, validation, and statistical analysis. This mapping from the Lagrangian trajectories to the Eulerian fields has also been adopted in other mesh-free methods, such as the field re-construction in SPH using the smoothing kernel. Notice, however,  that this particles-to-field mapping was used in SPH in a much more targeted way (than in how we use it here). Specifically, the SPH-mapping was tuned to approximate the spatial derivatives of the field quantities for approximating RHS of NS equations.   

Finally and third, if all the functions (represented by NNs) and constants are fixed by fitting Eqs.~(\ref{eqn:nnmodel}) from the Lagrangian DNS data,  the equations are closed.  Notice that this simple closure is in a contrast with what is built into the methodology of the SPH and of some other mesh-free methods. Indeed,  the SPH scheme also includes reconstruction of the pressure field from the density field, according to the equation of state, then followed by reconstruction of the velocity field from pressure by solving (in the incompressible case) the Poisson equation. Even though we do not have the density-to-pressure-to-velocity reconstruction step explicitly embedded into the L-LES Eqs.~(\ref{eqn:nnmodel}), we still reconstruct the fields from particles, because the reconstructed fields will be used in the following: (1) to build the Eulerian field-based loss function to learn parameters of the L-LES model, and (2) to test the results via statistical \emph{a-posteriori} analysis.

\section{From Lagrangian Trajectories to Eulerian Fields with Smoothing Kernel}
\label{sec:kernel}

Reconstruction of the fields (e.g. of velocity, density, and pressure) defined at an Eulerian grid, which is not changing in time, from the Lagrangian trajectories is one of the key elements of the SPH methodology where the temporally-evolving positions and velocities of the particles are considered as interpolation points for the fields evaluated at any other points of the domain. We follow the same basic approach and utilize the smoothing kernel method to reconstruct fields from the positions and velocities of the particles.  

Consider an exemplary scalar field, $A({\bm x})$, which may be one (of the three) components of the velocity field, the density field, or the pressure field. Image of the field convoluted against the smoothing kernel, $W_\theta({\bm x})$, parameterized by the vector of parameters, $\theta$ is 

\begin{equation}
\label{eqn:interpolation}
    \langle A(\bm{x}) \rangle = \int_\Omega A(\bm{x}') W_\theta(\bm{x}-\bm{x}')d\bm{x}'.
\end{equation}
The smoothing kernel, $W_\theta({\bm x})$, is assumed to be a continuous and sufficiently smooth function of compact support --- non-zero if $|\bm{x}-\bm{x}'|\leq h= c d$, where $c$ is a sufficiently large, $c > 1$, constant. This assumption, that the kernel is compactly supported, allows a more efficient implementation. Notice that the smoothing kernel also depends on the resolved scale, $d$. However, to simplify notations, here and below we will not show this dependence explicitly.   The smoothing kernel is also normalized
\begin{equation}
    \label{eqn:normalization}
\int_\Omega W_\theta(\bm{x}-\bm{x}') d\bm{x}' = 1,
\end{equation}
to guarantee the zeroth order consistency of the integral representation of continuum functions \cite{liu2010smoothed}. 

\begin{figure}[hbt]
    \centering
    \includegraphics[width=6in]{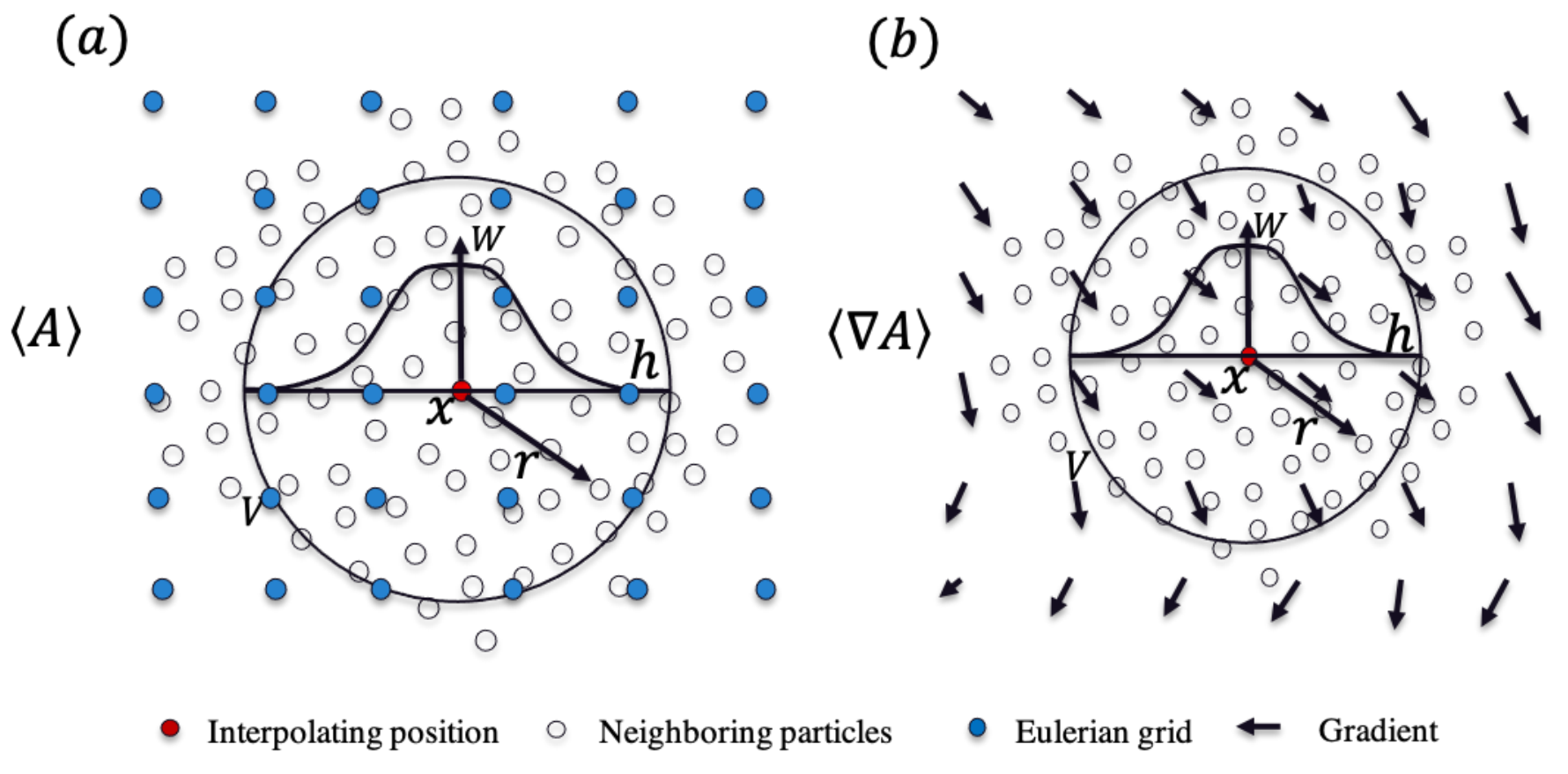}
    \caption{Schematic illustration of (a) the reconstruction of the image, $\langle A\rangle$, of the field, $A$, defined as a convolution of the kernel, $W_\theta(\cdot)$, with the field evaluated at the positions of the Lagrangian particles and (b) its gradient $\nabla \langle A\rangle$. {Width of the kernel, $h$, shown in the Figure, should be viewed as one of the components of the (tunable) vector of the kernel parameters, $\theta$.}
    }
    \label{fig:demo}
\end{figure}

Eq.~(\ref{eqn:interpolation}) allows us to reconstruct $\langle A \rangle$ at an field grid position $\bm{x}$ from the values of $A$ evaluated at the particle locations $\bm{x}_i$ 
\begin{equation}
\label{eqn:interpolationdiscrete}
\langle A(\bm{x}) \rangle = \sum_{j=1}^{N} \frac{m_j}{\rho_j} A(\bm{x}_j) W_\theta(\bm{x}-\bm{x}_j),
\end{equation}
where $m_j/\rho_j$ stands for a volume element associated with the particle $j$. A schematic illustration of the reconstruction of the field $\langle A(\bm{x}) \rangle$ is shown in Fig.~\ref{fig:demo} (a). We can also reconstruct the field of gradient of $A$, $\nabla\langle A(\bm{x}) \rangle$ (Fig. \ref{fig:demo} (b))using the following equation \cite{monaghan1985refined}:

\begin{subequations}
\begin{eqnarray}
  \langle \nabla A(\bm{x}) \rangle= \frac{1}{\rho_i} \l[ \sum_{j=1}^{N} m_j \l[ A(\bm{x}_j) -A(\bm{x})\r] \nabla_r W_{\theta}\r], \label{eqn:kernelderivative}
\end{eqnarray}
\end{subequations}

A classic SPH approach consists of postulating the form of the kernel function, $W_\theta(\cdot)$.  On the contrary, the approach of this manuscript (which was also developed in our earlier paper \cite{woodward2021physics}) aims at learning the kernel. Specifically, we model the parameterized kernel, subject to the finite-support and normalization constraints described above, using a neural network as follows

\begin{equation}
\label{eqn:pimlkernel}
    W_\theta({\bm r}) = \alpha_n \frac{\mathcal{NN}_\theta(r/h)
    }{1+\exp(C(r/h-1))}.
\end{equation}
$\mathcal{NN}_\theta(r)$ stands for a Neural Network (NN) with scalar input, $r=|{\bm r}|=|\bm{x}_{ij}|$, and scalar output; and $\alpha_n$ is the normalization coefficient enforcing Eq.~(\ref{eqn:normalization}). This specific choice of the functional form of $W_\theta(\cdot)$ ensures that the kernel is of compact support, by multiplying a sigmoid function, and that it is properly normalized. $C$ is a parameter that determines the sharpness of the decay to zero within the compact support domain.

The process of learning the smoothing kernel, $W_\theta(\cdot)$, i.e. training NN entering  (\ref{eqn:pimlkernel}), depends on two additional constructs described below. First, in Section \ref{sec:neighborhood}, we introduce the finite neighborhood list approximation, which is a technique that allows to reduce computational complexity of the convolution involving the smoothing kernel and thus to make the learning problem tractable. Then, in Section \ref{sec:loss-kernel} we discuss the loss function used to train the smoothing kernel. 

\subsection{Finite Neighborhood List Approximation}\label{sec:neighborhood}

Formally, the sum over particle pairs contributing the RHS of Eq.~(\ref{eqn:interpolationdiscrete}) evaluated at the $i$-th particle location, ${\bm x}={\bm x}_i$, contains $N$ terms.  However, for any $x$ there will be only $O(N_b)$ terms contributing significantly,  where $N_b$ is the number of particles that are $O(h)$ close to the $i$-th particle. To improve the efficiency of this evaluation, we prepare and update dynamically the finite neighborhood list for each particle in the system (i.e. the list consisting of the finite $O(1)$ number of neighbors for each particle), and then use it to truncate the number of terms contributing Eq.~(\ref{eqn:interpolationdiscrete}) to $O(1)$.

\subsection{Smoothing Kernel Loss Function} \label{sec:loss-kernel}

To describe the reconstruction of the velocity field in the Eulerian frame we introduce a constant, i.e. frozen in time, grid defined in terms of $N_E$ locations, ${\bm x}_a,\ a=1,\cdots, N_E$ which we index by the letters from the first three letters of the English alphabet, $a,b,c$ (that is in contrast with the Lagrangian particles, changing locations in time,  which are indexed by, $i,j,k=1,\cdots N$). 

To learn the smoothing kernel (SK) that maps Lagrangian particles to Eulerian fields, we build the SK Loss Function (LF), consisting of  three terms, 
\begin{gather}\label{eq:L-SK}
L_{SK} = c_v L_v + c_g L_g + c_n L_n.
\end{gather}

The first two terms in the SK-LF represent a mismatch between Lagrangian (particle) data and Eulerian (field) data for the velocities, $L_v$, and the velocity gradients, $L_g$, respectively: 

\begin{eqnarray}\label{eq:L_f}
L_v &=& \frac{1}{N_E} \sum_{a=1}^{N_E} \left|\langle \bm{v}(\bm{x}_{a}) \rangle - \bm{v}(\bm{x}_{a}) \right|^2,\\ 
    \langle \bm{v}(\bm{x}_{a}) \rangle &=& \int_\Omega \bm{v}(\bm{x}) W_\theta(\bm{x}_a-\bm{x}) d\bm{x} \approx \sum_{i=1}^{N_b} \bm{v}_{i} W_\theta(\bm{x}_{a}-\bm{x}_i) \Delta V_i,\\  \label{eq:L_g} 
    L_g &= &\frac{1}{N_E} \sum_{a=1}^{N_E} \sum_{\alpha =1}^{3}\left| \langle \nabla \bm{v}^{\alpha}(\bm{x}_{a})\rangle - \nabla \bm{v}^{\alpha}(\bm{x}_{a}) \right|^2,\\ \nonumber  
     \langle \nabla \bm{v}^{\alpha}(\bm{x}_{a})\rangle &=&
    \int_\Omega (\bm{v}^{\alpha}(\bm{x}) - \bm{v}^{\alpha}(\bm{x}_a)) \nabla_{r} W_\theta(\bm{x}_a-\bm{x}) d\bm{x} \nonumber \\ & \approx &
    \sum\limits_{i=1}^{N_b} \frac{m_i}{\Delta V_i} (\bm{v}^{\alpha}_i-\bm{v}^{\alpha}(\bm{x}_a)) \nabla_{r_{ai}} W_\theta(\bm{x}_{a}-\bm{x}_i) \frac{(\bm{x}_{a}-\bm{x}_i)}{r_{ai}},
\end{eqnarray}
where $ \Delta V_i = m_i/(\sum_{j=1}^{N} m_j W_\theta(\bm{x}_i-\bm{x}_j))$ is the volume element associated with the (Lagrangian) particle $i$ \cite{monaghan2012smoothed}; superscript $\alpha \in 1, 2, 3$ denotes the component of the velocity vector;  and values of ${\bm v}$, evaluated at the Lagrangian positions and locations of the Eulerian grid, are assumed taken from the \emph{ground truth} (GT) data extracted from properly filtered Lagrangian DNS (see Section \ref{sec:gtdata} for details).

There are, obviously, many possible choices one can make in designing the loss function. Specific choices of $L_v$ and  $L_g$ in Eqs.~(\ref{eq:L_f},\ref{eq:L_g},\ref{eq:L_i}) were guided by consideration of simplicity and also accepted practices of the SPH community.  In particular, we choose to work with the $l_2$ norm and we utilize the finite-particle asymmetric approximation for gradients from \cite{monaghan1992smoothed}, which is, according to \cite{liu2003smoothed,liu2006restoring}, helps to improve numerical accuracy.

The role of the third contribution to the SK-LF, $L_n$, is to enforce the SK normalization condition (\ref{eqn:normalization}), and thus learn the normalization coefficient $\alpha$ in Eq.~(\ref{eqn:pimlkernel}). Since the SK is modeled via NN, whose integral over $\Omega$ can not be computed analytically, we consider spherically symmetric SK, split the compact spherical domain of the integration described by Eq.~\ref{eqn:interpolation} into $N_r$ shells, of radii $r_k,\ k=0,1 ..., N_r$ and then approximate the remaining one-dimensional integral according to the trapezoidal rule, thus arriving at the following expression for the normalization enforcing component of $L_{SK}$:
\begin{gather}
\label{eq:L_i}
    L_n = (I-1)^2,\   I =\int_\Omega W_\theta(\bm{r}) d\bm{r}\approx  4\pi \Delta r \sum_{k=1}^{N_r-1}W_\theta(\bm{r}_k)r_k^2
\end{gather}

The Neural Network (NN) contributing the SK (\ref{eqn:pimlkernel}) is constructed using PyTorch \cite{NEURIPS2019_9015} open-source ML library. We build the neural network from four fully-connected layers with 20 nodes each, and with the hyperbolic tangent activation functions. The parameters (weights and biases of the NN layers) are randomly initialized using the Glorot normal initialization method. The parameters are updated by minimizing $L_{SK}$ using "Adam" (one of the most popular optimization methods). The initial learning rate is set to $10^{-3}$ and it is then reduced gradually to $10^{-6}$ throughout the training process. Training is terminated when both training and testing losses are saturated.

\section{Loss Function Enforcing Multi-Physics in L-LES}
\label{sec:piml}

To train the PIML L-LES model (\ref{eqn:nnmodel}) and verify that the model can reproduce realistic turbulence fields and statistics at the resolved scales, we need to introduce a Loss Function (LF) whose minimization embodies the training. There are a number of options (for the LF) to choose from. We present below some of these options and then rely on their linear combination in accomplishing the training task.

Since the L-LES model (\ref{eqn:nnmodel}) is principally Lagrangian, it is natural to consider a loss function stated in terms of particles. Therefore, let us, first of all, construct a trajectory-based LF, $L_t$, which compares \emph{prediction} of the L-LES model (\ref{eqn:nnmodel}) and respective GT data evaluated along the Lagrangian trajectories:

\begin{equation}\label{eq:L_t}
L_t = \frac{1}{N M} \sum_{i=1}^{N} \sum\limits_{m=1}^M \left| \frac{{\bm \phi}_i^{m+1}-{\bm \phi}_i^m}{\Delta}- \left(\begin{array}{l}\text{right-hand-side of Eq.~(\ref{eqn:nnmodel})}\\ \text{evaluated at }{\bm \phi}^{m}_i\end{array}\right)\right|^2
\end{equation}
were ${\bm \phi}^m_i$ stands for ${\bm \phi}_i(t)$ observed (as a Lagrangian part of the GT data) at the discretized moments of time $t_m$, $m=0,1,\cdots,M$; and we use a simple Euler method with time step $\Delta=t_{m+1}-t_m$ to find components of ${\bm \phi}$ integrating the L-LES model \eqref{eqn:nnmodel} in discrete time.

However,  we may also build a loss function based on a comparison of the observed and predicted values of the Eulerian field, thus resulting in
\begin{eqnarray}
\label{eqn:evolveFieldLoss}
L_{f} &=& \frac{1}{N_E M} \sum_{a=1}^{N_E} \sum_{m=1}^M \left|\langle\bm{\phi}_{a}^{m+1}\rangle - \bm{\phi}_{a}^{m+1}\right|^2, \\ \nonumber && \langle\bm{\phi}_{a}^{m+1}\rangle = \sum_{i=1}^{N} \frac{m_i}{\rho_i} \bm{\phi}^{m+1}_{i} W_\theta(\bm{x}_a^{m+1}-\bm{x}^{m+1}_i),\quad \bm{x}^{m+1}_{i} = \bm{x}^{m}_{i} + \Delta \bm{v}^{m+1}_{i},
\end{eqnarray}
where $({\bm \phi}_i^m, x_i^m, v_i^m| \forall i=1,\cdots,N,\ \forall m=1,\cdots,M)$ is available as the Lagrangian part of the GT data.

Finally, we construct the LF which aims to match the Lagrangian particle statistics -- predicted and observed. We choose to work with statistics of the single-particle Lagrangian velocities and accelerations -- which are the simplest possible Lagrangian characteristics testing the resolved scale. Therefore we aim to compare respective Probability Distribution Functions (PDFs) represented via histograms.

To build a histogram of a single-particle instantaneous velocity, $v_i^{\alpha}(t_n)$, 
we map the GT data $v^{\alpha}_i(t_m)$, aggregated over all the particles, $i=1,\cdots,N$, all the directions, $\alpha=1,2,3$, and all the available moments of time, $t_m,\ \forall m=1,\cdots,M$, to a histogram $P_v(\mu_k)$ build of $K$ bins, $k=1,\cdots,K,\ [\mu_k,\mu_{k+1}=\mu_k+w]$, each of size, $w$:
\begin{equation}\label{eq:hist}
    \forall k:\quad P_v(\mu_k)=\frac{1}{3 N M K}\sum\limits_{i=1}^N\sum_{\alpha=1,2,3}\sum\limits_{m=1}^M\mathds{1}\left(v_i^{\alpha}(t_m)\in[\mu_k,\mu_{k+1}]\right).
\end{equation}
Histogram of the particle acceleration, $P_{dv/dt}(\cdots)$ is built similarly.
Notice that the samples-to-histogram functions, $P_v(\mu_k)$ and $P_{dv/dt}(\mu_k)$, allow efficient representation via a Convolutional Neural Network (CNN), with some fixed weights and biases \cite{wang2016learnable},  which is designed specifically to enable efficient backpropagation over parameters (entering Eq.~(\ref{eq:hist}) via $v_i$). To compare statistics predicted by the model and statistics of the corresponding GT data we construct the Kullback-Leibler (KL) loss function for velocity (and similarly for acceleration):

\begin{equation}
L_{KL;v} 
= \sum_{k=1}^{m}P_{GT}(\mu_k) \text{log}\l[\frac{P_{GT;v}(\mu_k)}{P_{pred;v}(\mu_k)}\r]. 
\end{equation}

We combine the loss functions, representing different physical inputs into the structure of L-LES, in one expression,  
\begin{gather}\label{eq:L-LES-LF}
    L = c_t L_t + c_{f}L_{f} + c_{KL;v} L_{KL;v}+c_{KL;dv/dt} L_{KL;dv/dt},
\end{gather}
where $c_t$, $c_{f}$, $c_{KL;v}$ and $c_{KL;dv/dt}$ are constants which can be adjusted (e.g. during training for faster convergence).

\section{Ground Truth Data}
\label{sec:gtdata}

Our ``ground truth'' (GT) Lagrangian data are generated from the Eulerian Navier-Stokes (NS) Direct Numerical Simulation (DNS) of weakly compressible,  thus low Mach number, stationary Homogeneous Isotropic Turbulence (HIT). Our numerical implementation of the Eulerian DNS is 
on a $256^3$ mesh over the three-dimensional box, $\Omega=[0,2\pi]^3$. We use sixth-order compact finite differences for spatial discretization and the 4th-order Runge-Kutta scheme for time advancement, also imposing triply-periodic boundary conditions over the box. The velocity field is initialized with 3D Gaussian spectral density enforcing zero mean condition for all components. A large-scale quasi-solenoidal linear forcing term is applied to the simulation at wavenumber $\vert k\vert<2$ to prevent turbulence from decaying \cite{petersen2010}. The forcing method allows the specification of the Kolmogorov scale at the onset and ensures that it remains close to the specified value. The simulations presented here have $\eta /\Delta x =0.8$, where $\Delta x$ is the grid spacing. Compared to a standard (well-resolved) spectral simulation with $\eta k_{max}=1.5$, where $k_{max}$ is the maximum resolved wavenumber, which has $\eta/\Delta x=1.5/\pi$, the contraction factor is $\approx 0.6$ \cite{petersen2010,baltzer2020} and the maximum differentiation error at the grid (Nyquist) scale is less than $3.5\%$. Compared to a spectral method with $\eta k_{max}=1$, which has $\eta/\Delta x=1/\pi$, the contraction factor is $\approx 0.4$ \cite{petersen2010,baltzer2020} and the maximum differentiation error at the Nyquist scale is less than $0.2\%$. The initial temperature field is set to be uniform and the initial pressure field is calculated by solving the Poisson equation. More details about the numerical method and setup can be found in Refs. \cite{petersen2010,ryu2014}. The simulation is conducted until the turbulence becomes statistically stationary, which is verified based on the evolution of the kinetic energy and dissipation  \cite{petersen2010,ryu2014}.

Once a statistically-steady state of HIT is achieved, we remove the forcing term in the Eulerian DNS scheme and apply a Gaussian filter to the spatio-temporal Eulerian data to obtain the velocity field at the resolved scale,  $d$, and then inject in the filtered flow $64^3$ non-inertial Lagrangian fluid particles. We remove forcing to make sure that the flow, and consequently Lagrangian dynamics of passive particles, are not masked by non-universal details of the large-scale forcing. We use a Gaussian filter, which is commonly used in LES, with a filtering width of the order or larger than the scale $d$ that can be resolved for the $262144=64^3$ particles. (In the dimensionless units, where the energy-containing scale, $L$,  which is also the size of the box, is $L=2\pi$, the smallest $d$ we can resolve with this number of particles is  $\pi/32$, i.e. $64$ times smaller than the size of the domain.) 

The particles are placed in the computational domain at random, and then we follow trajectories of the passively advected particles for the time, $\tau$, which is of the order of (or longer) than the turbulence turnover time of an eddy of size comparable to the resolved scale, $d$, i.e. $\tau=O(d^{2/3}/\varepsilon^{1/3})$, where $\varepsilon$ is the estimate of the energy flux transferred downscale within the inertial range of turbulence \footnote{We remind that $d$ is bounded from above by the size of the box, i.e. $L=2\pi$ in the dimensionless units of our DNS setting, and from below by the Kolmogorov (viscous) scale, $\eta=O(\nu^{3/4}/\varepsilon^{1/4})$, where $\nu$ is the (kinematic) viscosity coefficient.}. 

In this work, we consider three turbulence cases for training and testing the model with comparable Reynolds numbers, $Re_\lambda \approx 80$, and the turbulent Mach number estimated as $M_t= 0.04$, $0.08$, and $0.16$, respectively. (See also Table \ref{table:use-cases}.) For fare comparison of different approaches to learning L-LES, we extract the filtered Eulerian velocity and velocity gradient snapshot at the same time as the Lagrangian particles. Particle positions and velocities $\{ \bm{x}_{i}, \bm{v}_{i}\}$ are then used to reconstruct velocities at the points of the Eulerian grid according to \eqref{eqn:interpolationdiscrete}. We follow Lagrangian trajectories for as long as the (largest) eddy turnover time  $\tau_{eddy}\sim (2\pi)^{2/3}/\varepsilon^{1/3}$, and utilize the data to compute and optimize the particle-based loss. The Eulerian fields are also recorded to compute and optimize the field-based loss.

\begin{table}
\centering
\caption{Three use cases considered in  training and validation of the L-LES model. \label{table:use-cases}}
\begin{tabular}{lrrr}
Case number & 1 & 2 & 3 \\
\hline
Turbulent Mach number $M_t$ & 0.08 & 0.16 & 0.04 \\
Taylor Reynolds number $Re_\lambda$ & 80 & 80& 80 \\
Kolmogorov timescale $t_\eta$ & 2.3 & 1.2 & 4.7 \\
Usage & training \& validation & validation & validation \\
\hline
\end{tabular}

\end{table}

\section{Results and Analyses}
\label{sec:results}

This Section presents results testing the quality of training of the Smoothing Kernel (SK), described by Eq.~(\ref{eqn:pimlkernel}),  and of the L-LES model, described by Eq.~(\ref{eqn:nnmodel}). We present the SK results first in Section \ref{sec:results-SK}. Here the focus is on analyzing the shape of the SK an testing the quality of the Eulerian field predictions for different choices of the coefficients in the SK-LF (\ref{eq:L-SK}). We then analyze in Section \ref{sec:results-L-LES} the L-LES model by investigating the generalization errors of the trained L-LES model and test the ability to reproduce turbulence statistics collected by simulating the trained L-LES model. 

\subsection{Smoothing Kernel}\label{sec:results-SK}

\begin{figure}[hbt]
    \centering
    \includegraphics[width=5in]{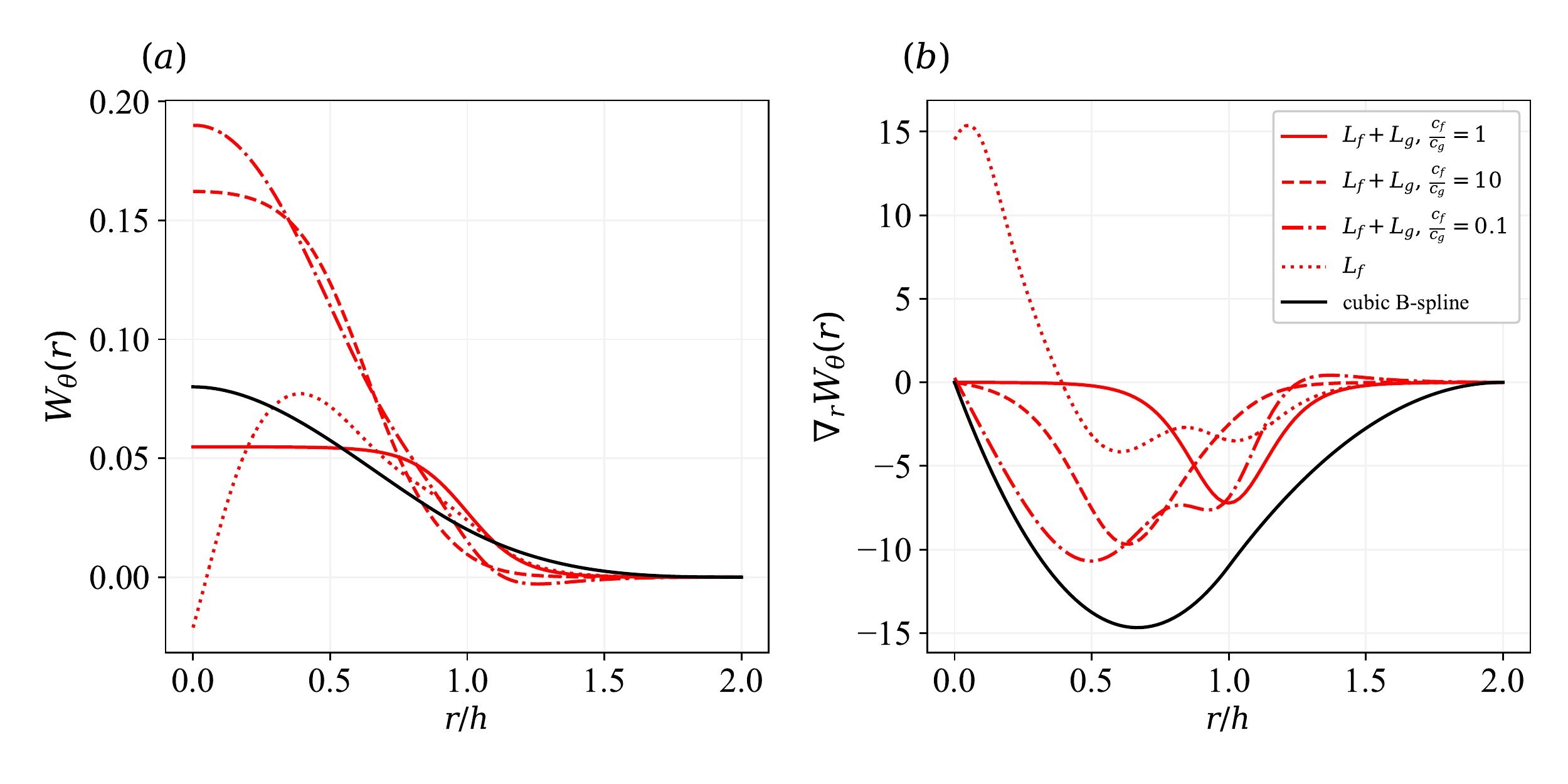}
    \caption{The learned smoothing kernels (a) and their derivatives to $r/h$ for different combinations of loss functions.
    }
    \label{fig:smoothingkernel}
\end{figure}

The learning of the SK (\ref{eqn:pimlkernel}) is discussed first because it is a prerequisite for learning the L-LES model (\ref{eqn:nnmodel}) and for the following statistical analyses. We experiment with training LF correspondent to different choices of the coefficients in Eq.~(\ref{eq:L-SK}). Figure (\ref{fig:smoothingkernel}) shows the resulting shape of SK and its derivative as a function of $r/h$ for two different choices of LF.  We also show as a reference a cubic spline shape of SK popular in the SPH studies \cite{monaghan1985refined}. We observe that the shape of the SK depends strongly on whether the gradient loss, $L_g$, is included in the LF. Without $L_g$, the learned SK does not show the decay property (that is the future empirically established, and thus widely adopted in the SPH studies \cite{liu2010smoothed}) and it shows a peak at $r/h \approx 0.4$. On the other hand, when $L_g$ is included in the LF, the anticipated decay of the SK with increase in $r$ is observed. This observation is significant because it \emph{explains} that decay of the SK is instrumental for an accurate prediction of the gradient field.

\begin{figure}[hbt]
    \centering
    \includegraphics[width=5in]{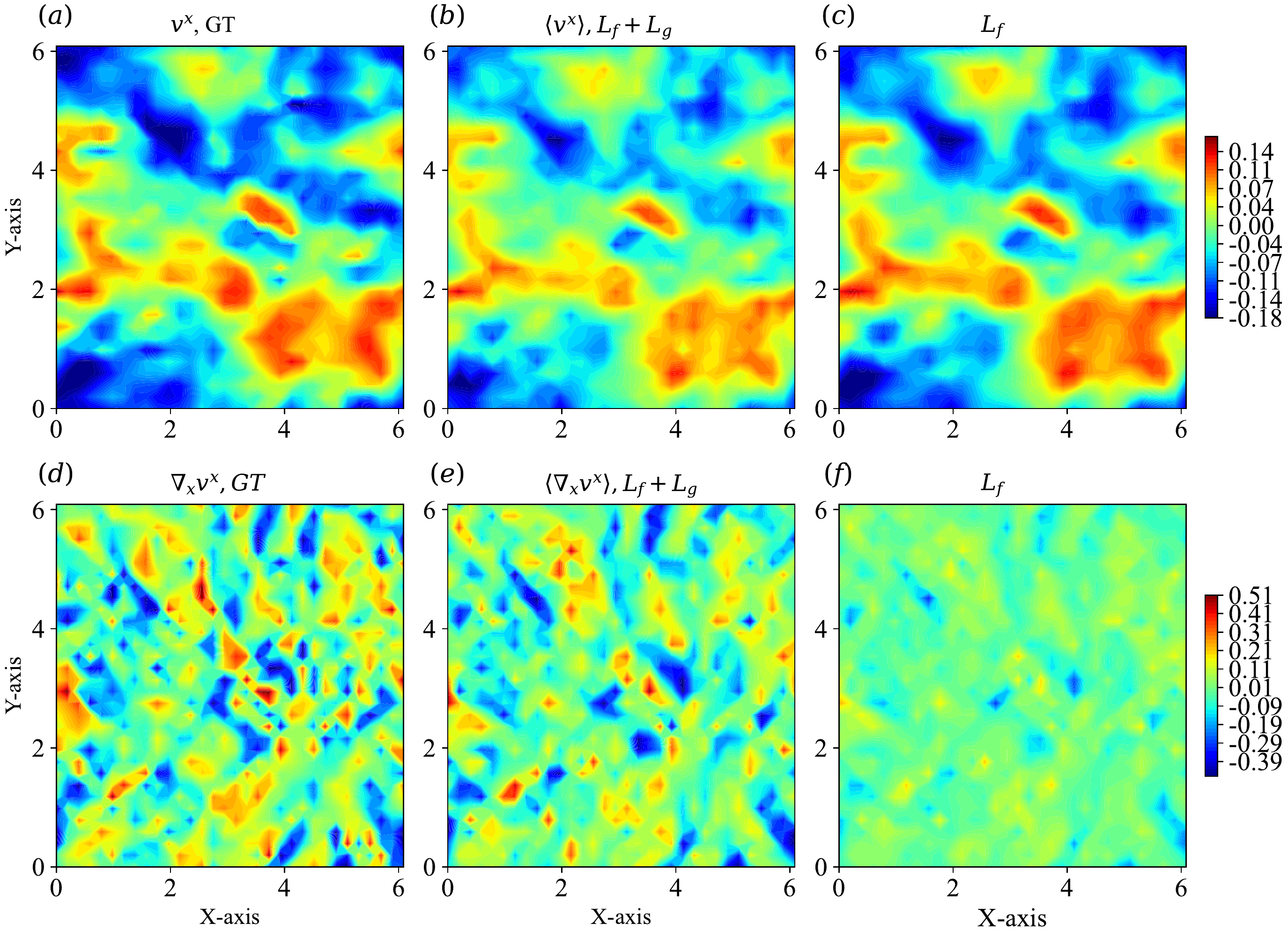}
    \caption{Two-dimensional contours of the three-dimensional velocity and velocity gradient field at $z=\pi$ reconstructed with the trained SK. Top row shows the $x$-component of the velocity vector and bottom row shows spatial derivative of $v^x$ in $x$-direction. The reconstructed field using different combinations of loss functions (b , c, e, f) are compared with ground truth DNS data (a,d). 
    }
    \label{fig:kernel_field}
\end{figure}

To examine qualitatively the reconstructed fields using the learned SK, we show in Fig. (\ref{fig:kernel_field}) the two-dimensional contours of the first velocity component $\langle v^{x}\rangle$  and its spatial derivative along the same direction, $\langle\nabla_{x} v^{x}\rangle$  using out-of-sample testing data. Comparing Fig.~\ref{fig:kernel_field} (a)-(c), we observe that the large-scale features of the flow can be reproduced using models trained with or without the gradient-based loss functions. On the other hand, the qualities of the reconstructed gradient field vary dependent on the LF. As shown in Fig.~\ref{fig:kernel_field} (d)-(e), the gradient field can not be correctly reproduced by models without accounting for the gradient term in the LF. In general, we observe (not shown in the figures) that even though the profile of the SK has a weak effect on the reconstruction quality of the field it influences very strongly the profile of higher derivatives. This observation emphasizes \emph{the significance of choosing correct SK}. A naive, i.e. not properly trained, choice of the smoothing kernel has a significant deterioration effect on predicting coarse-grained flows and their statistics.

\subsection{Training and Testing L-LES model}\label{sec:results-L-LES}

\subsubsection{Optimal Loss Function: A-priory Test}

In this Section we follow the discussion of Section \ref{sec:piml} and test the effects on the training results of the different L-LES model Loss Function contributions (different terms in Eq. (\ref{eq:L-LES-LF})). An exemplary  \emph{a-priori} test is set as follows: (a) We start with the trajectory-based LF, i.e. with all coefficients in Eq.~(\ref{eq:L-LES-LF}) but $c_t$ set to zero. This choice of the starting point is motivated by our empirical observation that the trajectory-based LF $L_t$ shows a faster training (decay rate of the LF). (b) When the decay rate saturates, we add to the LF all the other field and statistics-based contributions, $L_f$ and $L_{KL}$s. Notice that convergence of the field-based LF is significantly slower (than that observed initially with the trajectory-based loss). We attribute this observation to the additional layer of spatial averaging (summation over multiple particles) in  \eqref{eqn:evolveFieldLoss}. (c) The training is stopped once the decay of the combined LF is saturated.

\begin{figure}
    \centering
    \includegraphics[width=5in]{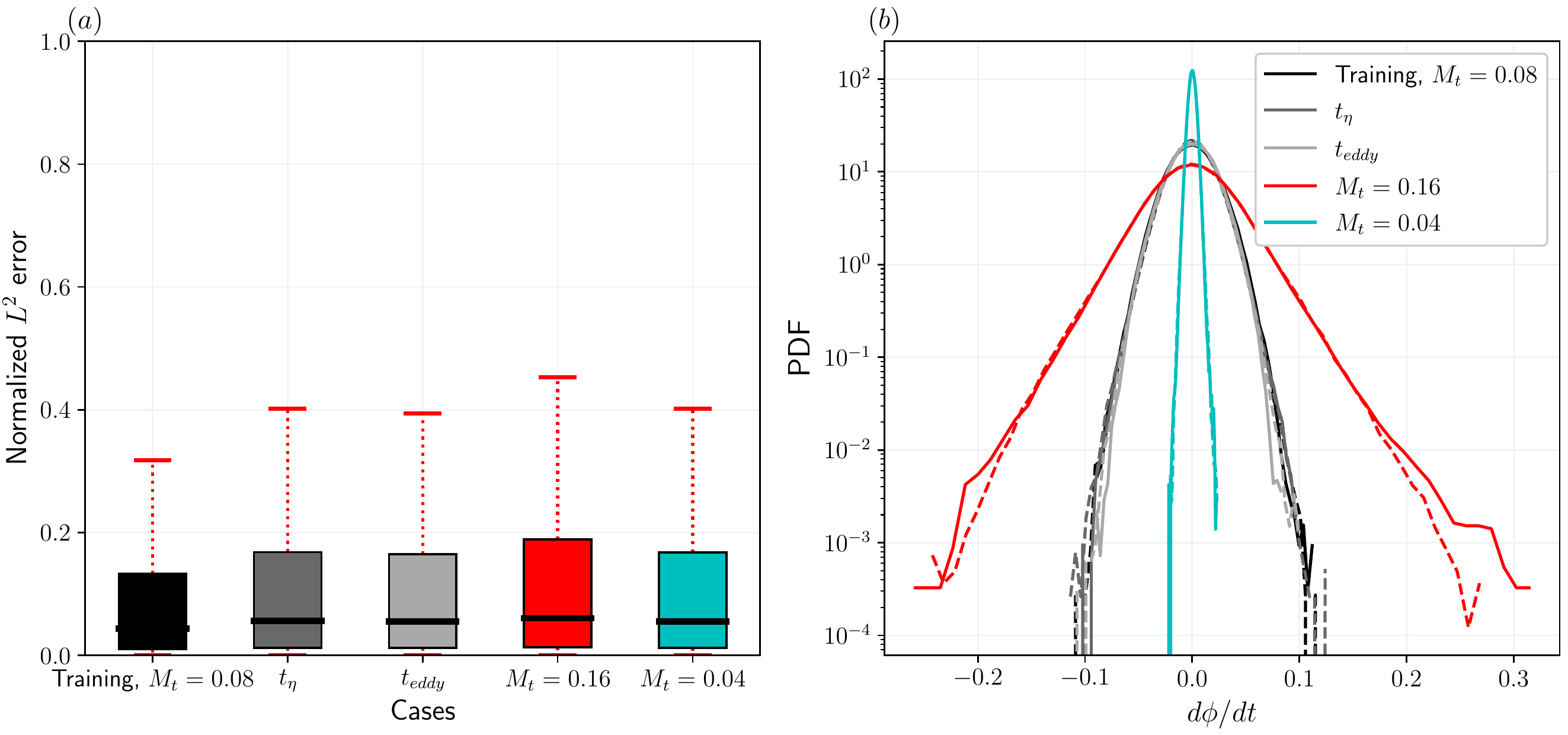}
    \caption{Performance on the trained L-LES model in interpolation and extrapolation errors over delayed intervals and different level of turbulent Mach number, respectively. We show (a) Normalized $L_2$ error; and (b) PDFs of the prediction.}
    \label{fig:error_generalization}
\end{figure}

We run a series of diagnostic tests of the L-LES model (\ref{eqn:nnmodel}). As custom in ML, we test the quality of the model training on the data not used in training. Specifically, we test if the trained model is capable to predict evolution from unseen sections of the flow field (interpolation) or flows with different characteristics (extrapolation). Fig.~\ref{fig:error_generalization} (a) shows  comparison of the normalized $L^2$ errors across different cases. 

To test the interpolation abilities of the model we apply it to the same turbulent model ($M_t \approx 0.08$) however predicting flows at times that are delayed to a different degree with respect to the flow segment used for training. We observe that prediction errors of the model, when applied with a delay, are very similar to the training errors. This indicates that the models do not over-fit and we can use it to interpolate in time. 

Our second test is more demanding. We test if the model allows extrapolation, i.e. generalizes well to new regimes with different $M_t$. Note that the L-LES model in the form of Eq. (\ref{eqn:nnmodel}) is dimensionless. After selecting the proper velocity, density, and acceleration scales based on the corresponding $M_t$, we are able to generalize the dynamics to different turbulence intensities and resolved scales. Results, testing different $M_t$, correspondent to different values of the resolved scale are also shown in Fig.~\ref{fig:error_generalization} (a). We observe that the model extrapolates very well to the regimes with different $M_t$s. 

Fig.~\ref{fig:error_generalization} (b) reports a comprehensive comparison of the acceleration statistics for the aforementioned interpolation and extrapolation experiments. We observe that the PDFs predicted by the trained L-LES model match the \emph{ground truth} data very well.

\subsubsection{Ultimate Test of the L-LES}

Once our choice of the Loss Function is validated (see preceding Subsection) we naturally turn to an in-depth, robust, and thus ultimate (we may also call it \emph{a-posteriori}) test of the Lagrangian features and prediction capabilities of the L-LES model.

Consider the following setup: 1) A random snapshot from the filtered \emph{Ground Truth} (GT) DNS simulation, which has not been used in training, is selected. In this ultimate test, we choose $M_t=0.16$, a turbulent Mach number never used in the training of the L-LES model. 2) Trained L-LES model \eqref{eqn:nnmodel} is initialized with $64^3$ particles dispersed (at random) at the locations $\bm{x}_i$ with their properties $\bm{\phi}_i$ interpolated using the respective Eulerian portion of the GT data.  3) The external force in the L-LES model (\eqref{eqn:nnmodel}) is chosen similar to the one implemented in the GT DNS Eulerian frame -- the energy injection rate is set to match the dissipation rate extracted from the decaying turbulence simulation. 4) We utilize the Varlet integration method to calculate the properties and positions of the Lagrangian particles  at the next step, $(\bm{x}^{n+1}_i,\ \bm{\phi}^{n+1}_i| i=1,\cdots,N)$ , given input from the previous step, $(\bm{x}^{n}_i,\ \bm{\phi}^{n}_i| i=1,\cdots,N)$. 5) This dynamic multi-particle simulation is run for  $t_{eddy}$, which is the turnover time of the turbulence energy containing scale. Then, we stop the simulation and perform dynamical and statistical analysis, some requiring interpolation of the Lagrangian particles to the Eulerian grid (according to the previously learned Smoothing Kernel).  

\begin{figure}
    \centering
    \includegraphics[width=6in]{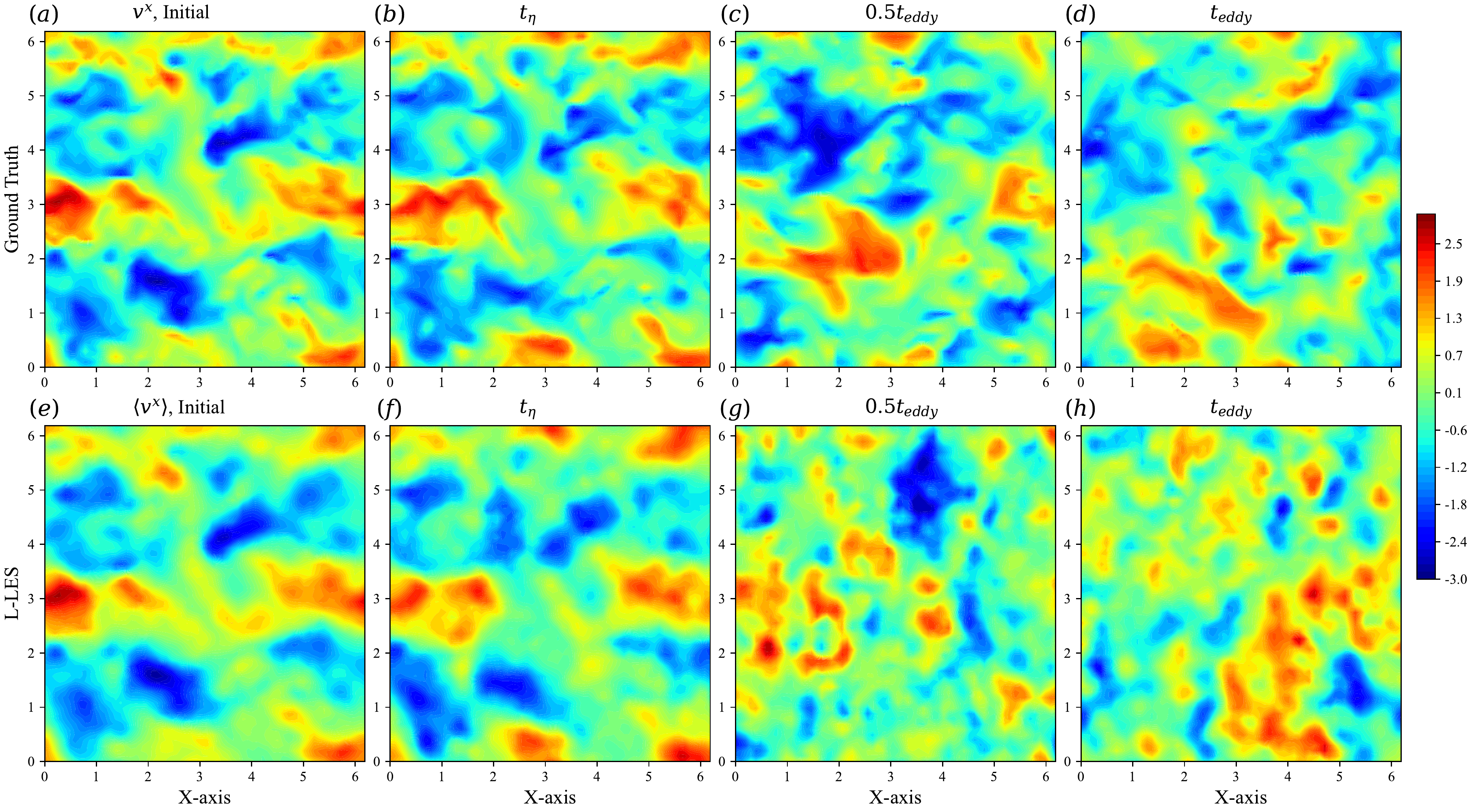}
    \caption{Two-dimensional contours of the three-dimensional velocity field at plane $z=\pi$ and turbulent Mach number $M_t = 0.16$. The top row shows the evolution of the $x$ component of the velocity vector from the initial snapshot (a) to different times scales, including Kolmogorov time scale (b) and eddy turnover time (c, d). The bottom row shows the reconstructed field by the learned SK using predictions from the L-LES model from the same initial condition.
    }
    \label{fig:contour_time}
\end{figure}

Fig.~(\ref{fig:contour_time}) shows the comparison of the two-dimensional contours at $z=\pi$ of predicted and GT velocity fields throughout the dynamic process just explained: the GT Eulerian fields at different moments of time are shown in the first row and the reconstructed fields are shown in the second row. We observe that early in the process,  e.g. at the times comparable to turnover time at the Kolmogorov scale, the contours match each other perfectly. Then, at the times corresponding to roughly a half of the eddy turnover time the Lagrangian simulations start to deviate from the exact trajectory of the GT field at the smaller scales. However, the large-scale structures of the two flows (predicted vs GT) still match each other reasonably well. When the time stamp approaches the eddy turnover time (right-most panel in Fig.~(\ref{fig:contour_time}) we start to observe a mismatch. In fact, the mismatch at the longest time stamp is not surprising for the following two reasons. First, our L-LES is a reduced-order model and as such it is not supposed to reproduce the dynamics of the truly chaotic system one-on-one. We do expect deviation growth in time eventually reaching the magnitude comparable to the flow itself (just as seen in Fig.~(\ref{fig:contour_time})). Second, the linear forcing is used in L-LES and is an approximation of the DNS forcing, but not exactly the same, therefore causing not very fast but still accumulation of the mismatch with time.

\begin{figure}
    \centering
    \includegraphics[width=4in]{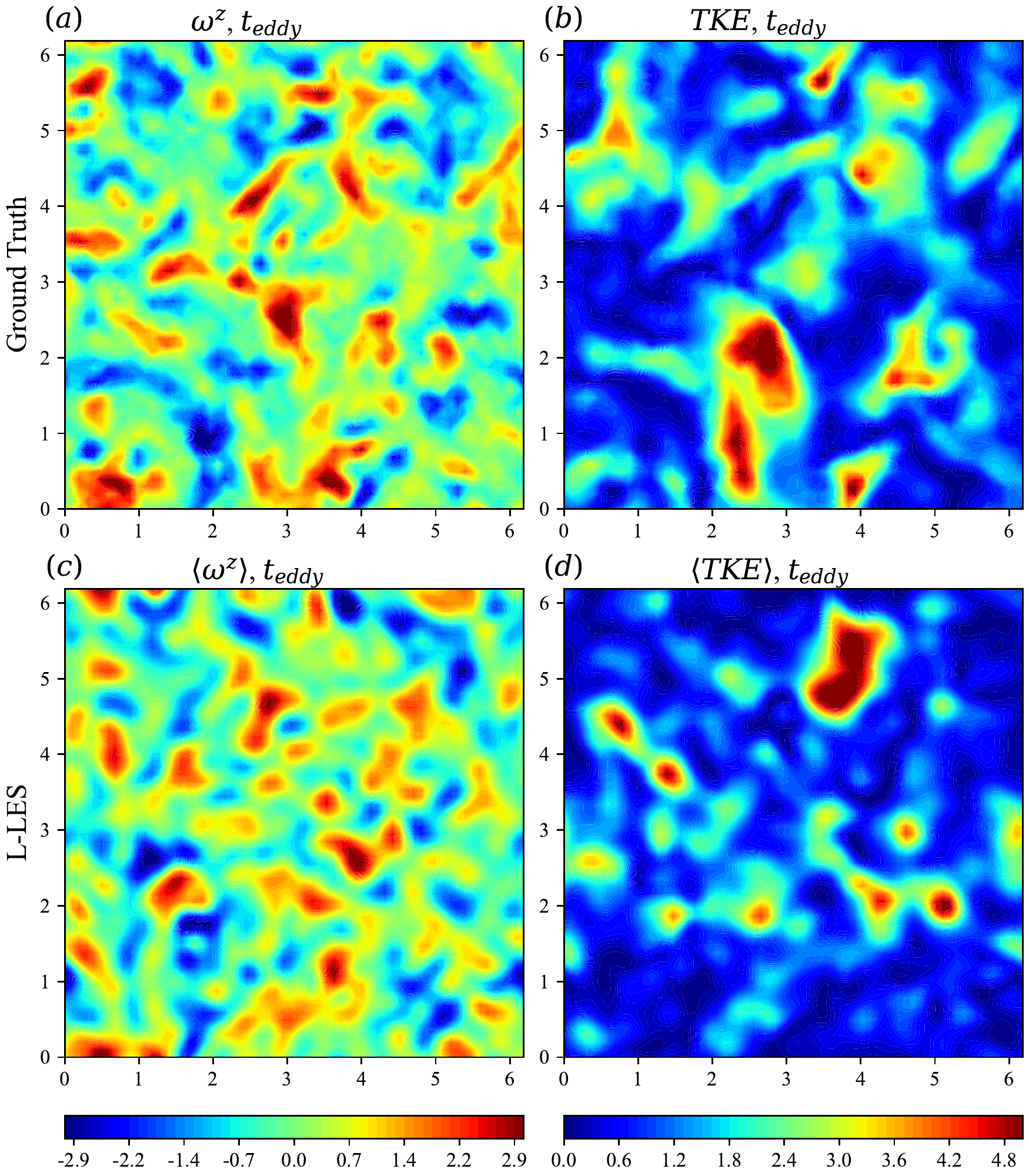}
    \caption{Two-dimensional contours of the three-dimensional vorticity (a,c) and turbulent kinetic energy (b,c) field at the plane $z = \pi$ and turbulent Mach number $M_t=0.16$. The top row shows the ground truth DNS field at the resolved scale $d$ at eddy turnover time and the bottom row shows the fields constructed by the learned smoothing kernel using predictions of the L-LES model from the same initial condition.}
    \label{fig:contour_property}
\end{figure}

Results of the statistical analysis of the ultimate test setting are shown in Fig.~ (\ref{fig:contour_property}).  Here, the two-dimensional contours of the vorticity and the turbulent kinetic energy (TKE) derived from the test are juxtaposed with these extracted from the GT data. We observe a reasonable qualitative agreement between the two.

\begin{figure}
    \centering
    \includegraphics[width=5in]{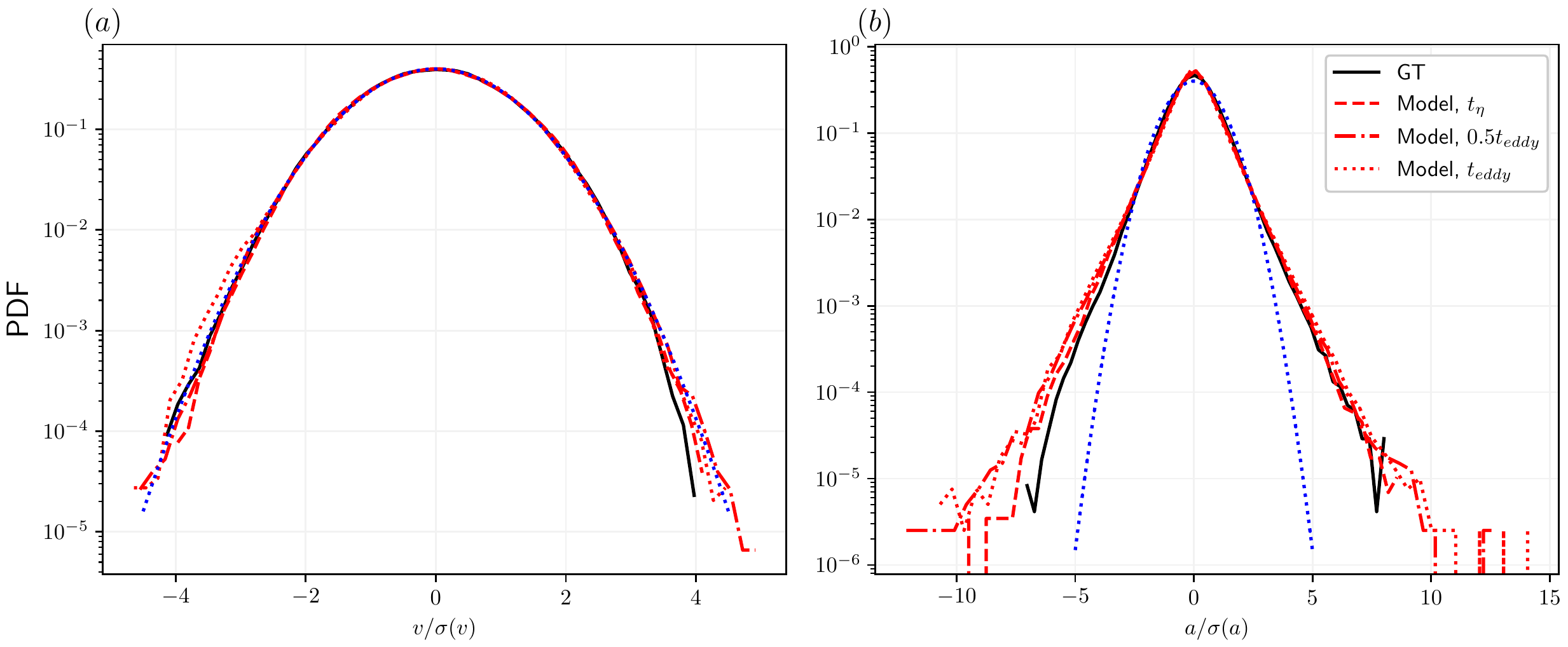}
    \caption{Normalized PDFs of the Lagrangian particle velocity and acceleration at different time scales. A reference Gaussian distribution (blue dotted line) is used for comparison.}
    \label{fig:pdf_u_a}
\end{figure}

Qualitative analysis of the velocity contours, reported in Fig.~(\ref{fig:contour_time}), shows the principal capability of the L-LES approach to predict flows fatefully.  However, such capabilities are also readily offered by the classic Eulerian LES. To demonstrate that L-LES can capture more physics, we turn now to a more quantitative analysis also emphasizing the Lagrangian perspective of the approach.

Fig.~(\ref{fig:pdf_u_a}) reports the results of the analysis capturing Lagrangian statistics of the turbulent flow. We show here the Probability Distribution Functions (PDFs) of the particle velocity and acceleration as captured by L-LES (vs ground truth) at the different time stamps (past the period where the GT data was available for training). 
We observe in Fig.~\ref{fig:pdf_u_a} (a) that the PDF of the particle velocity (coarse-grained at the resolved scale) is close to Gaussian. We also observe that L-LES predictions of the particle velocity at different time stamps match the respective GT data very well even if checked with a delay time (up to an eddy turnover time later). Fig.~\ref{fig:pdf_u_a} (b) shows similar comparison for the particle acceleration. Here again, we report a very good quantitative reproduction of the acceleration statistics by L-LES even when checked after evolving for one eddy turnover time.

\begin{figure}
    \centering
    \includegraphics[width=5in]{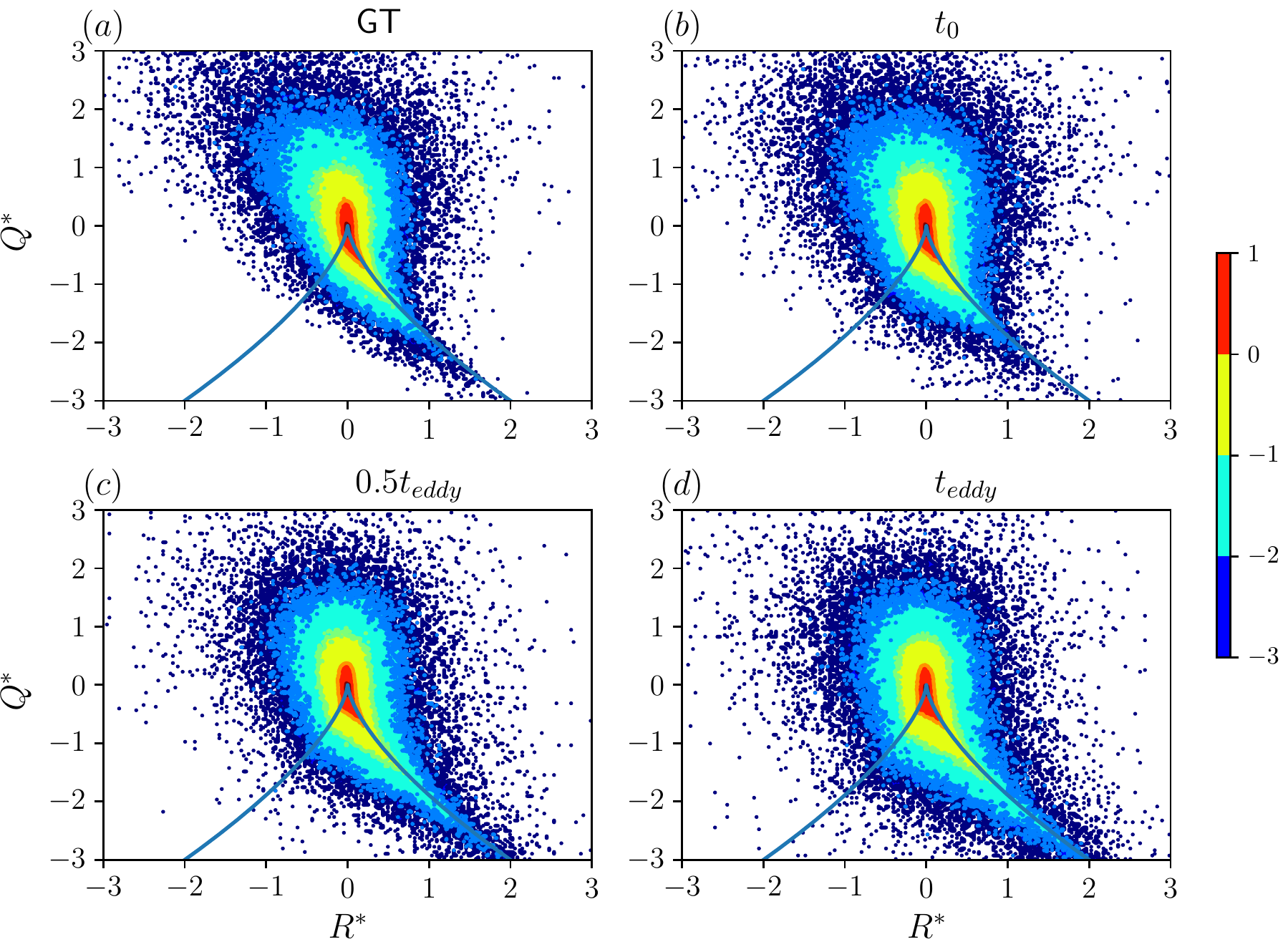}
    \caption{Statistical geometry of turbulence, as revealed by the joint PDF of the invariants of the coarse-grained velocity gradient ($Q$ and $R$). Subfigure (a) shows the ground truth (DNS). Subfigure (b) shows the PDF reconstructed for the moment equal to the beginning of the training window. Subfigures (c-d) show the PDF reconstructed by the L-LES model at the time stamps separated from the train window by $\tau_{eddy}/2$ and $\tau_{eddy}$, respectively. The VGT statistics is computed using the learned smoothing kernel (see text for details).}
    \label{fig:QR_time}
\end{figure}

\begin{figure}
    \centering
    \includegraphics[width=5in]{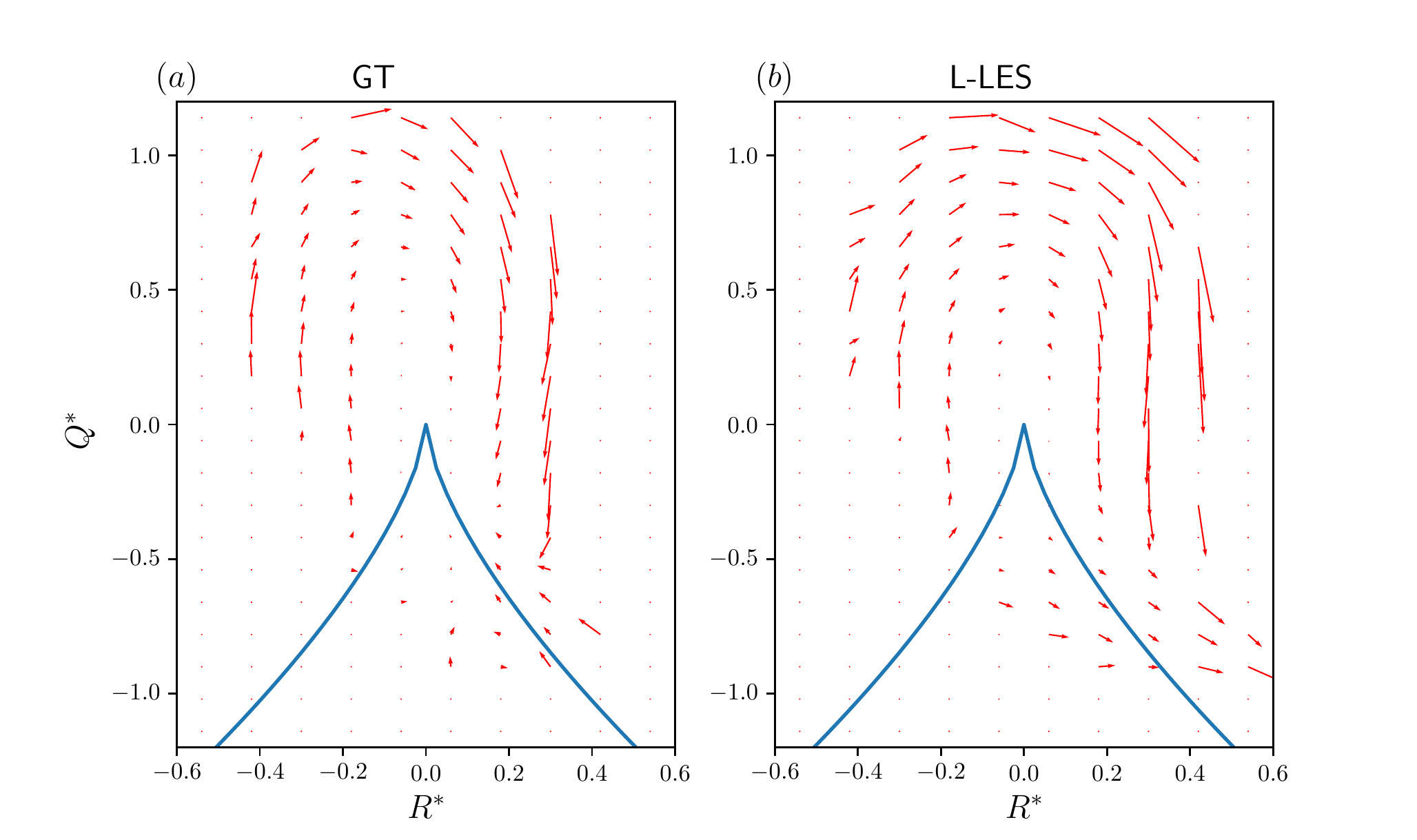}
    \caption{Vector field plot of the mean $(dQ/dt,dR/dt)$ conditioned to the values of the VGT invariants, $Q,R$, and averaged over Lagrangian particles, as extracted from the DNS data coarse-grained over, $d$ (a), and predicted by L-LES model (b).}
    \label{fig:QR_vector}
\end{figure}

Our statistical analysis so far was limited to relatively simple objects, e.g. velocity and acceleration statistics at the resolved scale.  Let us now turn to a more advanced diagnostics and test the so-called Q-R plane statistics of the coarse-grained velocity gradient \cite{Chertkov1999Lagrangian,tian2021physics}. Statistics in the $Q-R$ plane, where $Q$ and $R$ are second and third order invariants of the velocity gradient tensor coarse-grained at the resolved scale, $d$, is known to determine a variety of important characteristics of turbulence, such as the flow topology, deformation of material volume, energy cascade, and intermittency.

To test L-LES ability to reconstruct the joint PDF of $Q$ and $R$, we extract respective statistics following these steps: 1) We get ${\bm v}_i$ by running L-LES model (\ref{eqn:nnmodel}) and then compute $\nabla \bm{v}_i$ applying the Smoothing Kernel (\ref{eqn:interpolation},\ref{eqn:pimlkernel}). 2) We calculate the anisotropic part of the velocity gradient tensor  according to, $\nabla \bm{v}^{\ast}_i =\nabla \bm{v}_i - \frac{1}{3} \text{Tr}(\nabla \bm{v}_i )$. Then the second and third invariants of the anisotropic part of the velocity tensor are computed from

\begin{eqnarray}
Q_i &=& -\frac{1}{2} \text{Tr}(\nabla \bm{v}^{\ast}_i \times \nabla \bm{v}^{\ast}_i), \\
R_i &=& -\frac{1}{3} \text{Tr}(\nabla \bm{v}^{\ast}_i \times \nabla \bm{v}^{\ast}_i \times \nabla \bm{v}^{\ast}_i), \, \forall i \in 1,...,N
\end{eqnarray}
where $\times$ denotes matrix multiplication.

We calculate the second and third invariants for the particles at different time stamps from the L-LES predictions and aggregate them in a histogram/PDF, summing over particles. The results are shown in Fig.~(\ref{fig:QR_time}) in the form of the iso-lines of the joint PDF of the invariants. We observe the distinctive ``tear-drop" shape, expressing complex dynamics of turbulence, in all the subfigures of Fig.~(\ref{fig:QR_time}),  consistent with what is referred to in the literature as the statistical geometry of turbulence (see \cite{Chertkov1999Lagrangian} and references therein). Specifically, Fig.~\ref{fig:QR_time} (a) shows the $(Q,R)$ PDF from the ground truth (GT), DNS results. Fig.~\ref{fig:QR_time} (b) shows the $(Q,R)$ joint PDF reconstructed by L-LES at the moment of time corresponding to the initial condition of L-LES simulation. Figs.~\ref{fig:QR_time} (c,d) show the joint ($Q$,$R$) PDFs predicted by L-LES for the time stamp equal to half and one eddy turnover time (of the energy-containing scale), respectively. It is evident that the tear-drop shape of the joint ($Q$,$R$) PDF is reproduced by L-LES very well even at much later times, indicating the L-LES model is capable of reproducing detailed long-time statistics of Lagrangian turbulence.

The ability of L-LES to predict Lagrangian trajectories allows for the study of much more intimate features of turbulence, in particular statistics of the Velocity Gradient Tensor (VGT) as seen from the Lagrangian frame, i.e. PDF of the VGT accumulated over Lagrangian particles \cite{tian2021physics,tian2019jfm}. The Lagrangian dynamics of the VGT projected to the ($Q$,$R$) plane is shown in Fig.~(\ref{fig:QR_vector}). Specifically, we show in Fig.~(\ref{fig:QR_vector}) the mean (over Lagrangian particles) vector of the time derivatives of the VGT invariants, ($\frac{dR}{dt}, \frac{dQ}{dt}$), conditioned to the values of the invariants. We observe that the vector field prediction of the L-LES, shown in Fig.~\ref{fig:QR_vector} (b), agrees very well with the results extracted from the GT/DNS data coarse-grained at the resolved scale, $d$, shown in Fig.~\ref{fig:QR_vector} (a). The agreement is especially impressive given that we did not enforce the vector field matching in the loss function used to train the L-LES. In other words, the agreement between L-LES and DNS confirms that the former learns detailed Lagrangian features of turbulence from the latter accurately.

\begin{figure}
    \centering
    \includegraphics[width=5in]{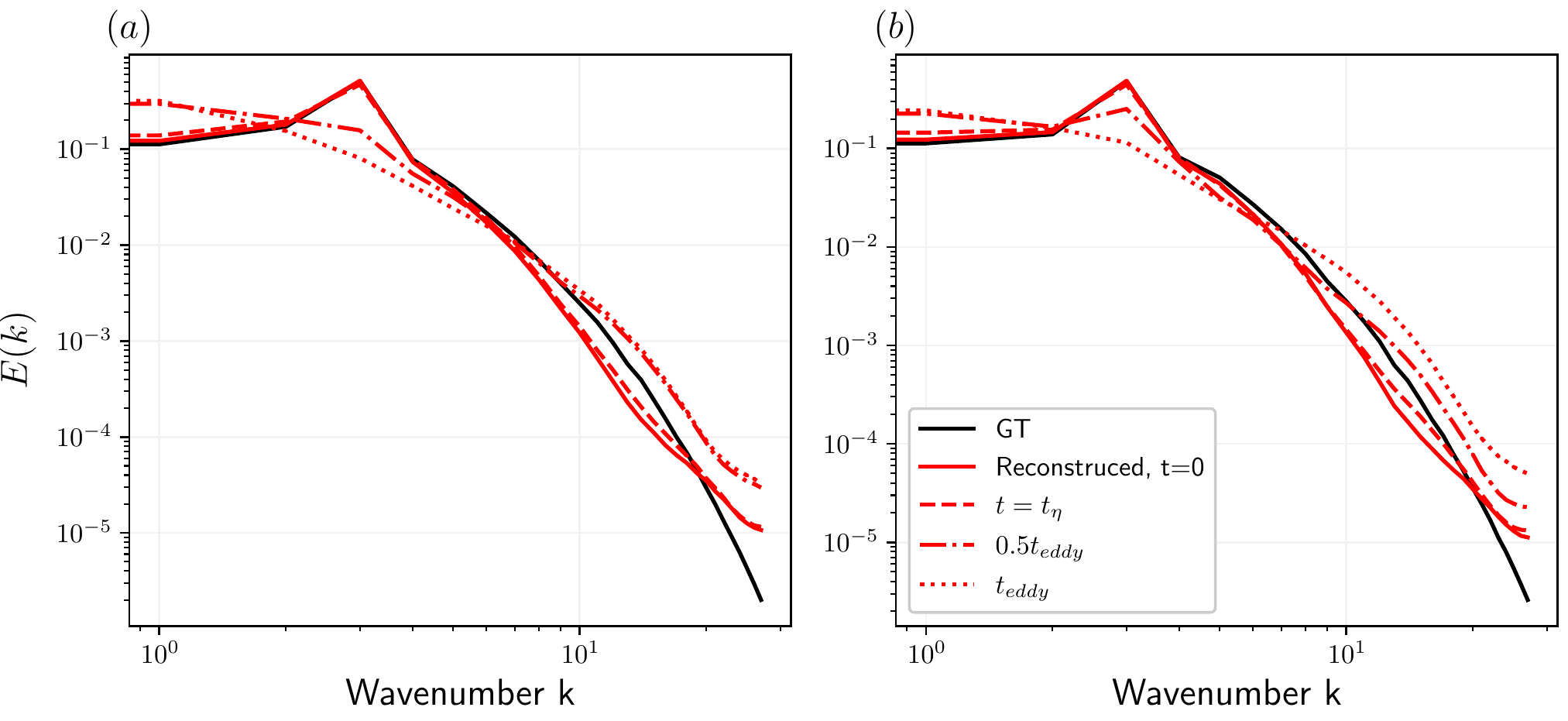}
    \caption{The energy spectrum of turbulence field as extracted from the Eulerian DNS/GT data vs reconstructed by L-LES at different time stamp. Subfigures (a) and (b) show comparisons at $M_t=0.16$ and $M_t=0.08$, respectively, while the training for both case is done at $M_t=0.08$.}
    \label{fig:spectrum}
\end{figure}

Fig.~\ref{fig:spectrum} shows evolution of the energy spectra for two different settings correspondent to $M_t = 0.08$ and $0.16$. Following our general logic, we train L-LES on the DNS/GT data correspondent to $M_t=0.08$ and then compare the energy spectra predictions at both $M_t = 0.08$ and $M_t=0.16$ with the results extracted from the DNS/GT. The L-LES energy spectra results are extracted from the Lagrangian (particles) data in two steps. First, we apply the Smoothing Kernel (SK) to Lagrangian (particle) data to reconstruct the snapshot of the velocity over the Eulerian grid, and then we compute the energy spectra (via standard Fourier transform), also truncating it at the wave vector correspondent to the resolved scale, $d$. We observe that at smaller scales, the energy spectra predicted by L-LES are slightly more intense than the energy spectrum extracted from the DNS/GT. We explain this overestimation of the turbulence intensity at the smaller scales by the fact that the inter-particle distance fluctuates, therefore leading to an additional energy stored at scales smaller than $d$. This is in contrast with the Eulerian DNS, which is filtered more precisely at the resolved scale, $d$. Looking at the large-scale part of the spectra we observe that as we increase the prediction time to $t_{eddy}$ the most energetic mode decreases. We attribute this to the fact that the forcing is not exactly the same in DNS and L-LES, i.e. the forcing term is applied in the spectral domain in DNS, but in L-LES it is applied in the spatial domain.

\section{Conclusions}
\label{sec:conclusions}

Representing turbulent flows with particles is advantageous because it potentially offers the ultimate \emph{model reduction} --- underlying physics becomes transparent and mathematics simpler. In the case of passive scalar turbulence, it allows to express $n$-th order correlation functions via dispersion of $n$ particles. In the case of "Burgulence", i.e. turbulence  described by Burgers equations, it allows to focus attention on the dynamics and interaction of shocks. 

This manuscript presents an attempt to carry over this Lagrangian logic to the much more challenging case of Navier-Stokes turbulence. We show how to build a reduced model of Homogeneous Isotropic Turbulence (HIT) in terms of a system of $N$ co-evolving Lagrangian markers/particles. Specifically, we claim fateful representation of HIT within the larger scale portion of the inertial range, which extends upscale from the resolved scale, $d$, to the energy-containing scale, $L$, in terms of at least $N$ particles, where $N=O( (L/d)^3)\gg 1$. 

We call this new approach \emph{Lagrangian Large Eddy Simulation} (L-LES), because it can be viewed as a Lagrangian version of the classic and highly popular LES approach, which was thus far solely Eulerian. 

One may as well argue that L-LES extends the growing list of particle-based approaches to modeling complex dynamics of fluids and materials. Insofar the \emph{Smooth Particle Hydrodynamics} (SPH),  which is arguably one of the most popular particle-based methods in fluid mechanics, is the L-LES closest ancestor. 

L-LES is also an advanced instance of the \emph{Physics Informed Machine Learning} (PIML) approach because it fits the ground truth data with the dynamic model which is sufficiently loose but also consistent with our current physical interpretation of the geometry and dynamics of HIT.  

L-LES, considered as a learning problem, is split in two parts. First, we learn/train the \emph{Smoothing Kernel} (SK), represented by Eqs.~(\ref{eqn:pimlkernel}),  which maps positions and velocities of the Lagrangian particles to the velocity and density fields evaluated over the Eulerian grid.  Then, we fit the L-LES model itself, represented by Eqs.~(\ref{eqn:nnmodel}). In both cases, the training is done on the DNS data and we rely on Neural Networks to provide the best fit to the degrees of freedom remaining after all the physical considerations and mathematical constraints, such as normalization of the SK, translational and rotational invariance, are taken into account. Physics also enters into the design of the \emph{Loss Function} (LF). 

The SK-LF is built from a combination of terms enforcing SK normalization and also matching the velocity field and the velocity gradient field with the SK applied to the respective ground truth (GT) Lagrangian data we obtained by advecting particles placed in the filtered (at the resolved scale) DNS.
Our experiments show that combining all three terms is critical for learning the SK. 

The L-LES Eqs.~(\ref{eqn:nnmodel}), describing Lagrangian dynamics of the probe particles at the resolved scale consistent with the filtered NS equations and thus accounting for the filtered stress tensor, sub-filter, viscous, and forcing contributions, are at the core of our modeling. We show that given the capability of a deep NN used to represent the remaining (and sufficiently large) freedom in the L-LES equations, the L-LES model is more inclusive than all previously considered particle-based models of turbulence, including SPH. Three types of LFs were introduced in order to train the L-LES model -- trajectory-based, field-based, and statistics-based. The model is trained on the Eulerian and Lagrangian GT data extracted from DNS simulations at turbulent Mach number $M_t=0.08$. Various \emph{a-priori} tests of the L-LES model all show that the model is capable to predict accurately multi-particle dynamics in unseen portions of the data at $M_t=0.08$ and it also generalizes (extrapolates) well to regimes with other values of $M_t$ within the weakly compressible limit. 

We also conducted \emph{a-posteriori} tests of advanced Lagrangian physics of turbulence where the model was not enforced directly. These tests/experiments included seeding Lagrangian particles into unseen (in training) parts of the GT data and collecting advanced statistics, e.g. on the \emph{energy spectra} and invariants of the filtered \emph{velocity gradients}. All of the tests have reported a quality agreement between the model and the GT.

\section*{Acknowledgement}{This work was performed under the auspices of DOE. Financial support comes from Los Alamos National Laboratory (LANL), Laboratory Directed Research and Development (LDRD) project "Machine Learning for Turbulence," 20180059DR. LANL, an affirmative action/equal opportunity employer, is managed by Triad National Security, LLC, for the National Nuclear Security Administration of the U.S. Department of Energy under contract 89233218CNA000001.}

\bibliographystyle{abbrv}
\bibliography{arXiv}
\end{document}